\documentclass[12pt]{article}
\usepackage{amsmath,amssymb,epsfig}
\usepackage{graphicx}

\title{%
     \begin{flushright}
     \vspace{-2cm}
     {\small KYUSHU-HET-48}
     \end{flushright}
     Ambiguities of theoretical parameters and CP/T violation in
     neutrino factories%
}

\author{%
   Masafumi Koike%
     \footnote{e-mail address: {\tt mkoike@post.kek.jp}}\\%
     {\footnotesize \it%
     Institute of Paticle and Nucear Studies,
     High Energy Accerelator Research Organization}\\
     {\footnotesize \it%
     1-1 Oho, Tsukuba, Ibaraki 305-0801, Japan%
     }
   \\
   Toshihiko Ota%
     \footnote{e-mail address: {\tt toshi@higgs.phys.kyushu-u.ac.jp}}\\%
     {\footnotesize \it%
     Department of Physics, Kyushu University,}\\
     {\footnotesize \it%
     Hakozaki, Higashi-ku, Fukuoka 812-8581, Japan
     }%
   \\
   and
   \\
   Joe Sato%
     \footnote{e-mail address: {\tt joe@rc.kyushu-u.ac.jp}}\\%
     {\footnotesize \it%
     Research Center for Higher Education, Kyushu University,}\\
     {\footnotesize \it%
     Ropponmatsu, Chuo-ku, Fukuoka 810-8560, Japan%
   }%
}
\date{}

\begin{document}

\maketitle

\begin{abstract}
    We study the sensitivity to the CP/T-violation search in the presence
    of ambiguities of the theoretical parameters.  Three generations
    of neutrinos are considered.  The parameters whose ambiguities are
    considered are the differences of the squared masses, the mixing
    angles, and the density of matter.
    
    We first consider the statistics that are sensitive to the genuine
    CP-violation effect originating from the imaginary coupling.  No
    ambiguity of the parameters is considered in this part.  It is
    argued that the widely-adopted usual statistics are not necessarily
    sensitive to the genuine CP-violation effect.  Two statistics that
    are sensitive to the imaginary coupling are proposed.  The
    qualitative difference between these statistics and the usual one 
    are discussed.
    
    Next we proceed to the case where the ambiguity of the parameters is
    present.  The sensitivity of the CP-violation search is greatly
    spoiled when the baseline length is longer than about one thousand
    kilometers, which turns out to be due to the ambiguity of the matter
    effect.  Thus the CP-violation search by use of CP conjugate
    channels turns out to require a low energy neutrino and short
    baseline length.  It is also shown that such a loss of sensitivity
    is avoided by using T-conjugate oscillation channels.
        
%
\end{abstract}

\section{Introduction}
The observation of the atmospheric neutrino anomaly by Super-Kamiokande
\cite{AtmSK} provided us with convincing evidence that neutrinos
have non-vanishing masses. There is another indication of neutrino
masses and mixings by the solar
neutrino deficit \cite{Ga1,Ga2,Kam,Cl,SolSK}.
%

These results give us the allowed region and excluded region for the
mixing angles and the mass square differences.  Let us now assume that
there are three flavors of neutrinos and denote the lepton mixing
matrix $U$, which relates the flavor eigenstates $\nu_{\alpha} (\alpha
= {\rm e}, \mu, \tau$) with the mass eigenstates $\nu_{i}$ with mass
$m_{i}$ $(i = 1, 2, 3)$ as $\nu_{\alpha} = \sum_{i=1}^{3} U_{\alpha i}
\nu_{i}$, by
\begin{eqnarray}
    U
    =
    \begin{pmatrix}
	c_{13} c_{12} & c_{13} s_{12} & s_{13}
	\\
	- c_{23} s_{12} - s_{23} s_{13} c_{12} {\rm e}^{{\rm i} \delta}
	&
	c_{23} c_{12} - s_{23} s_{13} s_{12} {\rm e}^{{\rm i} \delta}
	&
	s_{23} c_{13} {\rm e}^{{\rm i} \delta}
	\\
	s_{23} s_{12} - c_{23} s_{13} c_{12} {\rm e}^{{\rm i} \delta}
	&
	-s_{23} c_{12} - c_{23} s_{13} s_{12} {\rm e}^{{\rm i} \delta}
	&
	c_{23} c_{13} {\rm e}^{{\rm i} \delta}
    \end{pmatrix}.
    \label{eq:U-parametrization}
\end{eqnarray}
Here $c_{ij}$ and $s_{ij}$ stand for $\cos\theta_{ij}$ and
$\sin\theta_{ij}$, respectively.  The observations of the atmospheric
neutrino anomaly give us an allowed region for $\sin \theta_{23}$ and
the larger mass square difference ($\equiv\delta m^2_{31}$).  The
solar neutrino deficit provides allowed regions for $\sin \theta_{12}$
\cite{Fogli} and the smaller mass square difference ($\equiv \delta
m^2_{21}$).  On the other hand, the no-oscillation results of reactor
and accelerator experiments give us an exclusion region for
$\sin\theta_{13}$ (e.g. ref.\cite{Chooz}).  There is no constraint on
the CP-violating phase $\delta$.

The idea of neutrino factories with muon storage rings was proposed
\cite{Geer} to determine these mixing parameters (and the sign of
$\delta m^2_{31}$ in addition).  It attracted the interest of many
physicists
\cite{BGW,Golden,BGRW,BGRW2,GH,DFLR,FLPR,NuFact,BCR,KS,Yasuda}, and
the neutrino factories turned out to be a very promising candidate for
the next generation neutrino oscillation experiments.  We will be able to
observe neutrino oscillations even if $\sin \theta_{13}$ is as small as
0.01.  We will also be able to detect the CP-violation effects in such
experiments \cite{Cabibbo-CP,BWP}.
The possibility to observe CP violation through
long baseline neutrino oscillation experiments was discussed in Refs. 
\cite{Tanimoto,AJ,AKS,MN1,MN2,BGG} and many papers followed these works.

By what observation can we insist that we measure the CP violation? 
CP violation is characterized by the intrinsic imaginary part of a 
coupling in a Lagrangian.
The presence of an imaginary part of a coupling gives different
properties to particles and antiparticles, and it is observed to be
CP/T-violation effects.  Hence we need to discuss CP/T violation using
a quantity which is sensitive to the imaginary part of a coupling. 
We have to be very careful to construct such a quantity since
there is an indirect sensitivity to the CP violating phase (which
constitute not an imaginary part of couplings but a real part)
through a unitarity \cite{three}.
As for lepton CP/T violation in the long baseline neutrino oscillation
experiments, one of such quantities is the difference of event rate
between CP/T conjugate channels.  We must take care of matter effect \cite{MSW}
for CP conjugate channels since it gives difference to the event rate too.
Therefore we must take into consideration ambiguities of the parameters
as the matter effect can mimic the genuine CP violation partially.
We will show that the ambiguities of
parameters spoil the experimental sensitivity.%
\footnote{%
  It is also important to consider experimental systematic errors and
  backgrounds, but we do not consider them in this paper.  We assume
  that we can determine all the quantities such as particle energy. 
  The only error taken into account is statistical ones.%
}%

We formulate the treatment of the ambiguity of parameters within the
statistical method.  It is important to build a statistical quantity
that is sensitive to the imaginary part of coupling.  We propose a
proper statistics, and show explicitly that the sensitivity to CP
violation effect changes by taking the ambiguities into account.  To
this end we estimate how large exposure (proportional to number of
muons and detector size) is required to observe CP-violation effect by
the neutrino factory experiment.  The optimal experimental setup (muon
energy $E_\mu$ and baseline length $L$) is shown through such
considerations assuming three generation of neutrinos which account for
the solar neutrino deficit and the atmospheric neutrino anomaly.

This paper is organized as follows.  In section 2 we consider
statistical quantities which are proper to search for a genuine
CP-violation effect.  There we assume that the parameters such as
$\theta_{ij}$'s and $\delta m^{2}_{ij}$'s are known without
ambiguities.  We will discuss in section 3 the case where the
ambiguities of the parameters are taken into account. We present the
requirement on the number of muons and the mass of a detector to
observe genuine CP-violation effect through measurements of
CP-conjugate oscillation channels.  It is shown that the introduction
of the ambiguities of parameters greatly change the sensitivity to
the CP-violation search. 
In section 4 we investigate CP-violation search using T-conjugate 
oscillation channels.  Ambiguities of parameters are taken into 
account also in this section.  The sensitivity in this case is far 
better compared to the case using CP-conjugate channels, showing that 
this case is ideal (preferable) to search for CP-violation effect.
Finally a summary and the discussion is given in section 5.

                  %

\section{Direct observation of CP-violation effect}

Let us first discuss the physical quantity which characterizes the
presence of CP/T violation in the long baseline neutrino oscillation
experiments.  Such quantities must be sensitive to the imaginary part 
of the coupling.  We need to be particularly careful when we make 
use of CP-conjugate channels in the presence of the matter effect.


Let us first recall how the imaginary part of the lepton coupling gets
into the oscillation probabilities.  We use the oscillation in the
vacuum as a simplest example.  The oscillation probability from
$\nu_{\alpha}$ to $\nu_{\beta}$ in the vacuum is given by
\begin{eqnarray}
    & &
    P(\nu_{\alpha} \rightarrow \nu_{\beta}; E, L)
    =
    \sum_{i} 
    \left|
    U_{\beta i}
    {\rm e}^{-{\rm i} \delta m^{2}_{i} L / (2E)}
    U^{\ast}_{\alpha i}
    \right|^{2}
    \nonumber \\
    & = &
    \sum_{i,j}
    U^{\ast}_{\alpha i} U_{\alpha j} U_{\beta i} U^{\ast}_{\beta j}
    {\rm e}^{-{\rm i} \delta m^{2}_{ij} L / (2E)}
    \nonumber \\
    & = &
    \sum_{i, j}{\rm Re} \;
    U^{\ast}_{\alpha i} U_{\alpha j} U_{\beta i} U^{\ast}_{\beta j}
    \cos \frac{\delta m^{2}_{ij} L}{2E}
    +
    \sum_{i, j} {\rm Im} \;
    U^{\ast}_{\alpha i} U_{\alpha j} U_{\beta i} U^{\ast}_{\beta j}
    \sin \frac{\delta m^{2}_{ij} L}{2E},
    \label{eq:osc-P}
\end{eqnarray}
where $E$ and $L$ are the energy of neutrinos and the baseline length, 
respectively.  The unitarity of $U$ leads to 
\begin{equation}
    {\rm Im} \;
    U^{\ast}_{\alpha 1} U_{\alpha 2} U_{\beta 1} U^{\ast}_{\beta 2}
    =
    {\rm Im} \;
    U^{\ast}_{\alpha 2} U_{\alpha 3} U_{\beta 2} U^{\ast}_{\beta 3}
    =
    {\rm Im} \;
    U^{\ast}_{\alpha 1} U_{\alpha 3} U_{\beta 3} U^{\ast}_{\beta 1}
    \equiv J,
    \label{eq:U-unitarity}
\end{equation}
which allows us to write
\begin{equation}
    P(\nu_{\alpha} \rightarrow \nu_{\beta}; E, L)
    =
    \sum_{i, j}{\rm Re} \;
    U^{\ast}_{\alpha i} U_{\alpha j} U_{\beta i} U^{\ast}_{\beta j}
    \cos \frac{\delta m^{2}_{ij} L}{2E}
    +
    J \sum_{i, j}
    \sin \frac{\delta m^{2}_{ij} L}{2E}.
    \label{eq:osc-P-2}
\end{equation}
The Jarlskog parameter $J$ defined by eq.(\ref{eq:U-unitarity})
vanishes when all the $U_{ij}$'s are real, and the second term of
eq.(\ref{eq:osc-P-2}) also vanishes.  The existence of imaginary part
of lepton coupling gives non-vanishing $J$, and thus we need to observe
the quantity which is sensitive to $J$ (including its sign) to search directly for
CP-violation effect.  Note that this statement is independent of the
parametrization of $U$ such as in eq.(\ref{eq:U-parametrization}).

Now we consider the quantity which is sensitive to the Jarlskog
parameter $J$ in presence of matter on the baseline.  We put some
assumptions for simplicity to discuss this point.  Suppose that we
have same initial energy spectra for neutrinos and antineutrinos. 
Also we suppose that the antineutrinos have about twice larger data
size so that the expected event numbers for neutrinos and
antineutrinos are equal in no-oscillation case.  Then the oscillation
event numbers of neutrinos $N(\nu_{\alpha} \rightarrow \nu_{\beta})$
and that of antineutrinos $N(\bar \nu_{\alpha} \rightarrow \bar
\nu_{\beta})$ are expected to be equal in vacuum if CP is conserved. 
The two event numbers are in practice different due to CP violation
(if any) and matter on the baseline.  The difference of the two event
numbers,
\begin{equation}
    \Delta N (\delta)
    \equiv
    N(\nu_{\alpha} \rightarrow \nu_{\beta}; \delta)
    -
    N(\bar \nu_{\alpha} \rightarrow \bar \nu_{\beta}; \delta)
    \label{eq:Delta-N}
\end{equation}
is intuitively sensitive to the genuine CP violation.  Here the
CP-violating angle $\delta$ is explicitly written.  This quantity does
not vanish due to matter effect, even in absence of genuine CP
violation.  We thus consider
\begin{equation}
    \Delta N(\delta) - \Delta N(\delta_{0}),
    \label{eq:DeltaN-DeltaN0}
\end{equation}
where $\delta = \delta_{0} \in \{ 0, \pi \}$ corresponds to the CP
conserving case.

We stress here that the quantity
\begin{equation}
    N(\nu_{\alpha} \rightarrow \nu_{\beta}; \delta)
    -
    N(\nu_{\alpha} \rightarrow \nu_{\beta}; \delta_{0})
    \label{eq:param-fitter}
\end{equation}
is not necessarily sensitive to the genuine CP violation.  To
compare eqs.(\ref{eq:Delta-N}) and (\ref{eq:param-fitter}), let us
consider the oscillation probability $P(\nu_{\alpha} \rightarrow
\nu_{\beta}; \delta)$, which is related to $N(\nu_{\alpha} \rightarrow
\nu_{\beta}; \delta)$ roughly by
\begin{equation}
    N(\nu_{\alpha} \rightarrow \nu_{\beta}; \delta)
    \sim
    \frac{E^{3}}{L^{2}} P(\nu_{\alpha} \rightarrow \nu_{\beta}; \delta).
    \label{eq:N-P-relation}
\end{equation}
It can be shown in the high energy limit%
\footnote{%
  The limit is valid when $E_{\nu} \gtrsim \delta \tilde m^{2}_{31} 
  L / 4$, where $\delta \tilde m^{2}_{31} \equiv \sqrt{(\delta 
  m^{2}_{31} \cos 2 \theta_{13} - a)^{2} + (\delta m^{2}_{31} \sin 2 
  \theta_{13})^{2}}.$
}%
\cite{Ota}
\begin{eqnarray}
    & &
    P (\nu_{\mu} \rightarrow \nu_{\rm e}; \delta, a)
    \nonumber \\
    & = &
    \left(
    \frac{\delta m_{31}^{2} L}{4E}
    \right)^{2}
    \left[
    B + \frac{\delta m^{2}_{21}}{\delta m^{2}_{31}}
    ( j \cos \delta - 2 B \sin^{2} \theta_{12} )
    \right]
    \left[
    1 - \frac{1}{3} \left( \frac{a L}{4E} \right)^{2}
    \right]
    \nonumber \\
    & + &
    \left(
    \frac{\delta m_{31}^{2} L}{4E}
    \right)^{3}
    \frac{a L}{4E}
    \times
    \nonumber \\
    & &
    \left\{
    \frac{2}{3} B \cos 2 \theta_{13} 
    +
    \frac{\delta m^{2}_{21}}{\delta m^{2}_{31}}
    \left[
    \frac{1}{3}
    j \cos \delta 
    \left( 2 \cos 2 \theta_{13} - 1 \right)
    -
    2
    B \cos 2 \theta_{13} \sin^{2} \theta_{12}
    \right]
    \right\}
    \nonumber \\
    & - &
    \left(
    \frac{\delta m_{31}^{2} L}{4E}
    \right)^{3}
    \frac{\delta m^{2}_{21}}{\delta m^{2}_{31}}
    j \sin \delta
    \nonumber \\
    & + &
    {\cal O} \left( \frac{1}{E^{4}} \right),
    \label{eq:hiE-prob-nu}
\end{eqnarray}
where
\begin{eqnarray}
    j 
    &\equiv&
    \sin 2 \theta_{12} \sin 2 \theta_{23}
    \sin 2 \theta_{13} \cos \theta_{13},
    \label{eq:j-def}
    \\
    B
    &\equiv&
    \left| U_{\rm e3} \right|^{2}
    \left| U_{\mu 3} \right|^{2}
    =
    \sin^{2} \theta_{23} \sin^{2} 2 \theta_{13},
    \label{eq:B-def}
    \\
    a
    &\equiv&
    2 \sqrt{2} G_{\rm F} n_{\rm e} E,
    \label{eq:a-def}
\end{eqnarray}
and $n_{\rm e}$ in eq.(\ref{eq:a-def}) is the average electron number
density in the matter on the baseline.  Recall that the Jarlskog
parameter $J$ defined by eq.(\ref{eq:U-unitarity}) is expressed under
the parametrization eq.(\ref{eq:U-parametrization}) by
\begin{equation}
    J = \sin 2\theta_{12} \sin 2\theta_{23} \sin 2\theta_{13} 
    \cos \theta_{13} \sin \delta,
    \label{eq:J-def}
\end{equation}
which is related to $j$ in eq.(\ref{eq:j-def}) by
\begin{equation}
    J = j \sin \delta.
    \label{eq:Jj-rel}
\end{equation}
Thus the third term of eq.(\ref{eq:hiE-prob-nu}) is the contribution
from genuine CP-violation effect.  Note again that this statement is
also independent of the parametrization.

We obtain from eq.({\ref{eq:hiE-prob-nu}})
\begin{eqnarray}
    & &
    P (\nu_{\mu} \rightarrow \nu_{\rm e}; \delta, a)
    -
    P (\nu_{\mu} \rightarrow \nu_{\rm e}; \delta_{0}, a)
    \nonumber \\
    & = &
    \left(
    \frac{\delta m_{31}^{2} L}{4E}
    \right)^{2}
    \frac{\delta m^{2}_{21}}{\delta m^{2}_{31}}
    j ( \cos \delta \mp 1 )
    \left[
    1 - \frac{1}{3} \left( \frac{a L}{4E} \right)^{2}
    \right]
    + {\cal O} \left( \frac{1}{E^{3}} \right)
    \label{eq:hiE-paramFitProb}
\end{eqnarray}
($-$ sign for $\delta_{0} = 0$ and $+$ for $\delta_{0} = \pi$) and
\begin{eqnarray}
    & &
    P (\nu_{\mu} \rightarrow \nu_{\rm e}; \delta, a)
    -
    P (\bar \nu_{\mu} \rightarrow \bar \nu_{\rm e}; \delta, a)
    \nonumber \\
    & = &
    P (\nu_{\mu} \rightarrow \nu_{\rm e}; \delta, a)
    -
    P (\nu_{\mu} \rightarrow \nu_{\rm e}; -\delta, -a)
    \nonumber \\
    & = &
    2 \left(
    \frac{\delta m_{31}^{2} L}{4E}
    \right)^{3}
    \times
   \nonumber \\
   & &
    \left(
    \frac{a L}{4E}
    \left\{
    \frac{2}{3} B \cos 2 \theta_{13}
    +
    \frac{\delta m^{2}_{21}}{\delta m^{2}_{31}}
    \left[
    \frac{1}{3} j \cos \delta
    \left( 2 \cos 2 \theta_{13} - 1 \right)
    - 2 B \cos 2 \theta_{13} \sin^{2} \theta_{12}  
    \right]
    \right\}
    \right.
    \nonumber \\
    & &
    \left.
    -
    \frac{\delta m^{2}_{21}}{\delta m^{2}_{31}}
    j \sin \delta
    \right)
    \nonumber \\
    & + & 
    {\cal O} \left( \frac{1}{E^{4}} \right).
    \label{eq:hiE-deltaProb}
\end{eqnarray}
We can observe in eq.(\ref{eq:hiE-paramFitProb}) that the leading term
of $P (\nu_{\mu} \rightarrow \nu_{\rm e}; \delta, a) - P (\nu_{\mu}
\rightarrow \nu_{\rm e}; \delta_{0}, a)$ does not contain the genuine
CP-violation term.  This leading term can be canceled by taking the
difference between the probabilities of neutrinos and antineutrinos. 
Equation (\ref{eq:hiE-deltaProb}) is indeed sensitive to the $j \sin
\delta$ term, though it contains an unavoidable matter effect term in
addition.

Our viewpoint is that the CP-violation search must be carried out
directly by observing the contribution of the $j \sin \delta$ term in
eq.(\ref{eq:hiE-prob-nu}), which originates from the {\it imaginary}
part of the coupling as in eqs.(\ref{eq:U-unitarity}),
(\ref{eq:J-def}) and (\ref{eq:Jj-rel}).  This term is not the leading
term at least in high energy region, but it can be picked up as a
leading term by taking the difference between neutrinos and
antineutrinos as in eq.(\ref{eq:hiE-deltaProb}).  Applying this
consideration also to event rates, we regard that
eq.(\ref{eq:DeltaN-DeltaN0}) is a better quantity than
eq.(\ref{eq:param-fitter}) to pursue the possibility of direct 
CP-violation search.  An analysis using eq.(\ref{eq:param-fitter}) is an
usual parameter fitting method.  It does not take into careful
consideration whether or not $N(\nu_{\alpha} \rightarrow \nu_{\beta};
\delta)$ is sensitive to imaginary part of the coupling.  Even if one
obtains a result that $N(\delta) \neq N(\delta_{0})$ in an experiment
which is sensitive only to the real part, there remains a possibility
to build a certain Lagrangian with totally {\it real} coupling which
can reproduce $N(\delta)$.  In this respect $N(\delta) \neq
N(\delta_{0})$ cannot be the definite clue of the presence of CP
violation.

Let us further exemplify the difference between
eqs.(\ref{eq:DeltaN-DeltaN0}) and (\ref{eq:param-fitter}).  We
consider a following toy setup of an experiment.  A source of neutrino
beam is $N_{\mu}$ muons which decay into neutrinos at a muon ring. 
The neutrinos extracted from the ring are detected at a detector if
their energy $E_{\nu}$ is larger than a threshold energy $E_{\rm th}$. 
The detector has mass $M_{\rm detector}$ and contains $N_{\rm target}$
target atoms, which are related as
\begin{equation}
    N_{\rm target}
    =
    6.02 \times 10^{34} \frac{M_{\rm detector}}{\rm [100kt]}.
    \label{eq:target2detector}
\end{equation}
We assume the neutrino-nucleon cross section $\sigma$ is proportional
to neutrino energy as
\begin{equation}
    \sigma = \sigma_{0} E_{\nu},
    \label{eq:sigma}
\end{equation}
where
\begin{equation}
    \sigma_{0}
    =
    \left\{
    \begin{array}{ll}
        0.67 \times 10^{-38} {\rm cm^{2} / GeV} & \mbox{for neutrinos,} \\
        0.34 \times 10^{-38} {\rm cm^{2} / GeV} & \mbox{for antineutrinos.} \\
    \end{array}
    \right.
    \label{eq:def-sigma0}
\end{equation}
The expected number of appearance events in the energy bin $E_{j-1} <
E_{\nu} < E_{j}$ ($j = 1, 2, \ldots, n$) is then given by
\begin{equation}
    N_{j} (\nu_{\alpha} \rightarrow \nu_{\beta}; \delta)
    \equiv
    \frac{ N_{\mu} N_{\rm target} \sigma_{0} }{ \pi m_{\mu}^{2} }
    \frac{ E_{\mu}^{2} }{ L^{2} } 
    \int_{ E_{j-1} }^{ E_j }
    E_{\nu}
    f_{\nu_{\alpha}} (E_{\nu})
    P(\nu_{\alpha} \rightarrow \nu_{\beta}; \delta)
    \frac{ {\rm d} E_{\nu} }{ E_{\mu} },
    \label{eq:app-event-number}
\end{equation}
where $m_{\mu}$ is the muon mass, and $f_{\nu_{\alpha}}(E_{\nu})$ is the
neutrino flux that is concretely given by eqs.(\ref{eq:nue-spectrum}) and 
(\ref{eq:numu-spectrum}).  
We define
\begin{equation}
    C
    \equiv
    \frac{\sigma_{0}}{\pi m_{\mu}^{2}}
    \frac{N_{\rm target}}{M_{\rm detector}}
     \label{eq:defN0}
\end{equation}
and
\begin{equation}
    R_{j}(\nu_{\alpha} \rightarrow \nu_{\beta}; \delta)
    \equiv
    \int_{E_{j-1}}^{E_j}
    E_{\nu}
    f_{\nu_{\alpha}} (E_{\nu})
    P(\nu_{\alpha} \rightarrow \nu_{\beta}; E_{\nu}, \delta)
    \frac{{\rm d} E_{\nu}}{E_{\mu}},
    \label{eq:defAveP}
\end{equation}
so that
\begin{eqnarray}
    & &
    N_{j} (\nu_{\alpha} \rightarrow \nu_{\beta}; \delta)
    =
    N_{\rm \mu} M_{\rm detector} \frac{E_{\mu}^{2}}{L^{2}}
    C R_{j}(\nu_{\alpha} \rightarrow \nu_{\beta}; \delta)
    \nonumber \\
    & = &
    \left\{
    \begin{array}{l}
	1.14 \\
	0.58
    \end{array}
    \right\}
    \times 10^{3}
    \frac{N_{\mu}}{[10^{21}]} \frac{M_{\rm detector}}{\rm [100 kt]}
    \left( \frac{E_{\mu} / {\rm [GeV]} }{ L / {\rm [1000 km]} } \right)^{2}
    \times
    R_{j}(\nu_{\alpha} \rightarrow \nu_{\beta}; \delta)    
    \label{eq:Nj-short}
\end{eqnarray}
(1.14 for neutrinos and 0.58 for antineutrinos).  A quantity $N_{\mu}
M_{\rm detector}$ is normalized in unit of $\rm [10^{21} \cdot 100kt]$
in eq.(\ref{eq:Nj-short}).  The value of unity in this unit is a quite
optimistic one compared to the presently discussed values
\cite{BGRW2}.  The requirement on $N_{\mu} M_{\rm detector}$ in this
normalization must be about unity or less so that we can observe the
CP-violation effect experimentally.

The widely adopted $\chi^{2}$ statistical quantity based
on eq.(\ref{eq:param-fitter}) is defined by
\begin{eqnarray}
    \chi_{1}^{2} (\delta_{0})
    & \equiv &
    \sum_{j=1}^{n}
    \frac
    {[ N_{j}(\delta) - N_{j}(\delta_{0}) ]^{2}}
    {N_{j}(\delta)}
    +
    \sum_{j=1}^{n}
    \frac
    {[ \bar N_{j}(\delta) - \bar N_{j}(\delta_{0}) ]^{2}}
    {\bar N_{j}(\delta)}
    \nonumber \\
    & = &
    N_{\mu} M_{\rm detector}
    \frac{E_{\mu}^{2}}{L^{2}}
    C
    \left\{
    \sum_{j=1}^{n}
    \frac
    {[ R_{j}(\delta) - R_{j}(\delta_{0}) ]^{2}}
    {R_{j}(\delta)}
    +
    \sum_{j=1}^{n}
    \frac
    {[ \bar R_{j}(\delta) - \bar R_{j}(\delta_{0}) ]^{2}}
    {\bar R_{j}(\delta)}
    \right\},
    \label{eq:chi1-0-def}
\end{eqnarray}
where
\begin{equation}
    N_{j} (\delta)
    \equiv
    N_{j} (\nu_{\alpha} \rightarrow \nu_{\beta}; \delta), \quad
    \bar N_{j} (\delta)
    \equiv
    N_{j}(\bar\nu_{\alpha} \rightarrow \bar\nu_{\beta}; \delta),
    \label{eq:Nj&Nj-bar}
\end{equation}
\begin{equation}
    R_{j} (\delta)
    \equiv
    R_{j} (\nu_{\alpha} \rightarrow \nu_{\beta}; \delta), \quad
    \bar R_{j} (\delta)
    \equiv
    R_{j}(\bar\nu_{\alpha} \rightarrow \bar\nu_{\beta}; \delta)
    \label{eq:Rj&Rj-bar}
\end{equation}
and $n$ is the number of bins.  Similarly we define based on 
eq.(\ref{eq:DeltaN-DeltaN0}) as\footnote{This quantity depends on a certain model 
through subtraction of the matter effect. 
However, if we can observe this asymmetry significantly, then we would
be able to conclude that there is a genuine CP-violation effect in the
real Lagrangian, even if the real theory is not the model that we assume.
}
\begin{equation}
    \chi_{2}^{2} (\delta_{0})
    \equiv
    \sum_{j=1}^{n}
    \frac
    {\left[
    \Delta N_{j}(\delta) - \Delta N_{j}(\delta_{0}) \right]^{2}
    }{
    N_{j}(\delta) + \bar N_{j}(\delta)
    }
    =
    N_{\mu} M_{\rm detector}
    \frac{E_{\mu}^{2}}{L^{2}}
    C
    \sum_{j=1}^{n}
    \frac
    {
    \left[
    \Delta
    R_{j}(\delta) - \Delta R_{j}(\delta_{0})
    \right]^{2} }
    { R_{j}(\delta) + \bar R_{j}(\delta) },
    \label{eq:chi2-0-def}
\end{equation}
where
\begin{equation}
    \Delta R_{j}(\delta) \equiv R_{j}(\delta) - \bar R_{j}(\delta).
    \label{eq:Delta-Rj}
\end{equation}
We need both $\delta \neq 0$ and $\delta \neq \pi$ to ascertain that 
CP violation is present.  We thus define
\begin{eqnarray}
    \chi_{1}^{2}
    & \equiv &
    \min_{\delta_{0} \in \{  0, \pi\} }
    \chi_{1}^{2}(\delta_{0}),
    \label{eq:chi1-def} \\
    \chi_{2}^{2}
    & \equiv &
    \min_{\delta_{0} \in \{  0, \pi\} }
   \chi_{2}^{2}(\delta_{0}).
    \label{eq:chi2-def}
\end{eqnarray}
and require
\begin{eqnarray}
    \chi^{2}_{1}
    & > &
    \chi^{2}_{90\%}(2 n),
    \label{eq:chi1-cond} \\
    \chi^{2}_{2}
    & > &
    \chi^{2}_{90\%}(n)
    \label{eq:chi2-cond}
\end{eqnarray}
to claim that the CP-violation effect is observable at 90\% confidence
level in the method with $n$ energy bins; 
more details on statistics is found in the Appendix.  
Equations (\ref{eq:chi1-cond}) and (\ref{eq:chi2-cond}) is equivalent to
\begin{eqnarray}
    N_{\mu} M_{\rm detector}
    & > &
    (N_{\mu} M_{\rm detector})_{\rm min; 1, 90\%}
    \nonumber \\
    & \equiv &
    \frac{1}{C}
    \frac{L^{2}}{E_{\mu}^{2}}
    \frac{ \chi^{2}_{90\%}(2 n) }
    {\displaystyle
    \min_{\delta_{0} \in \{ 0, \pi \}} 
    \left\{
    \sum_{j=1}^{n}
    \frac
    {[ R_{j}(\delta) - R_{j}(\delta_{0}) ]^{2}}
    {R_{j}(\delta)}
    +
    \sum_{j=1}^{n}
    \frac
    {[ \bar R_{j}(\delta) - \bar R_{j}(\delta_{0}) ]^{2}}
    {\bar R_{j}(\delta)}
    \right\}
    }
    \label{eq:C1-con}
\end{eqnarray}
and
\begin{eqnarray}
    N_{\mu} M_{\rm detector}
    & > &
    (N_{\mu} M_{\rm detector})_{\rm min; 2, 90\%}
    \nonumber \\
    & \equiv &
    \frac{1}{C}
    \frac{L^{2}}{E_{\mu}^{2}}
    \frac{ \chi^{2}_{90\%}(n) }
    {\displaystyle
    \min_{\delta_{0} \in \{ 0, \pi \}}
    \left\{
    \sum_{j=1}^{n}
    \frac
    {
    \left[
    \Delta
    R_{j}(\delta) - \Delta R_{j}(\delta_{0})
    \right]^{2} }
    { R_{j}(\delta) + \bar R_{j}(\delta) }
    \right\}
    },
    \label{eq:C2-cond}
\end{eqnarray}
respectively.

%
\begin{figure}
 \unitlength = 1cm
 \begin{picture}(15,9)
  \unitlength = 1mm
  \centerline{
   \epsfysize = 9cm
   \epsfbox{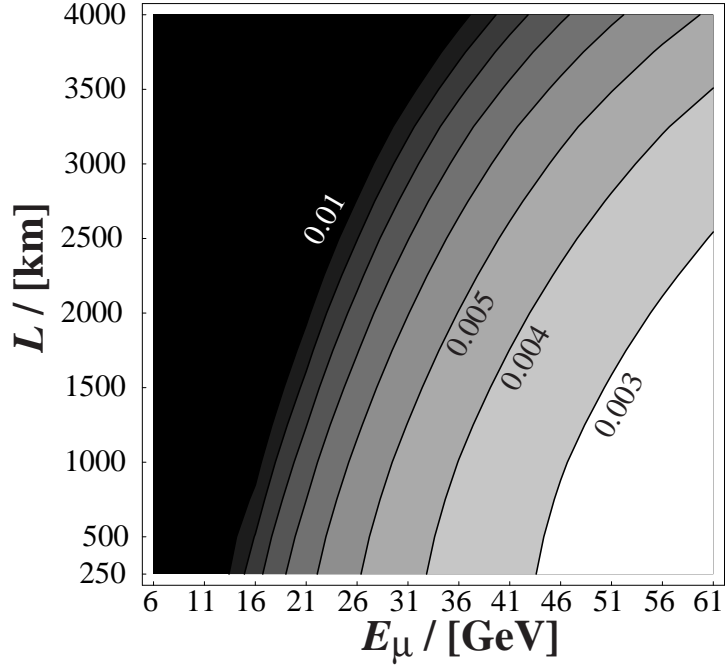}
  }
 \end{picture}
 \caption{%
 A contour plot of the required data size to observe the genuine
 CP-violation effect.  A quantity $(N_{\mu} M_{\rm detector})_{\rm
 min; 1, 90\%}$ defined in eq.(\ref{eq:C1-cond}) is plotted in
 unit of $\rm [10^{21} \cdot 100kt]$ as a function of muon energy and
 baseline length.  Smaller value of this value means the higher
 sensitivity.  Here $E_{\rm th} = 1 {\rm GeV}$, and the case of
 $\delta = \pi / 2$ is presented.  Other parameters are taken as shown
 in eqs.(\ref{eq:graph-angles}) and (\ref{eq:graph-m^{2}}).  The
 smaller value of $(N_{\mu} M_{\rm detector})_{\rm min; 1, 90\%}$
 means the higher sensitivity of the CP-violation search.  The use of
 $\chi^{2}_{1}$ leads to the higher
 sensitivity as $E_{\mu}$ gets larger.%
 }
\label{fig:chi1}
\end{figure}
%
%
\begin{figure}
 \unitlength = 1cm
 \begin{picture}(15,9)
  \unitlength = 1mm
  \centerline{
   \epsfysize = 9cm
   \epsfbox{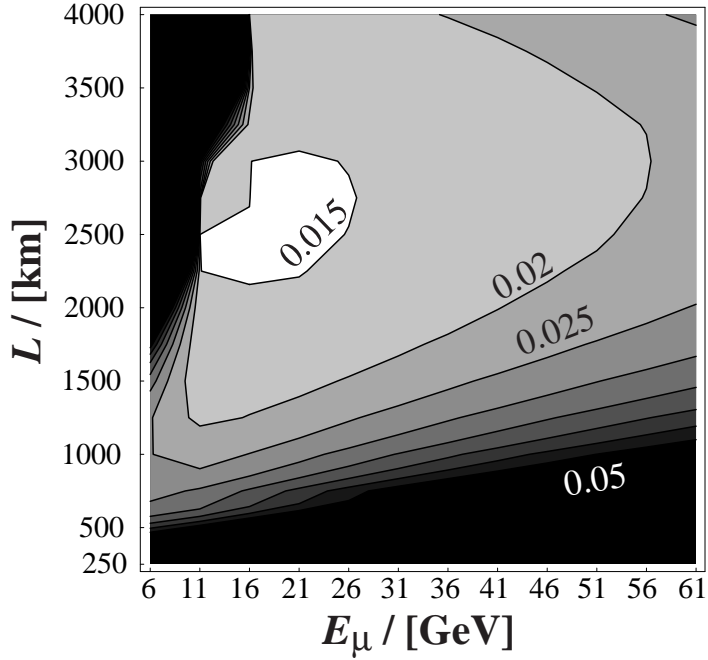}
  }
 \end{picture}
\caption{%
  A contour plot of $(N_{\mu} M_{\rm detector})_{\rm min; 2, 90\%}$ in
  unit of $\rm [10^{21} \cdot 100kt]$.  The parameters are the same
  as in Fig.\ref{fig:chi1}.  A larger data sample is needed compared
  to Fig.\ref{fig:chi1}.  Optimum muon energy and baseline length
  makes a sweet spot in the graph.%
}
\label{fig:chi2}
\end{figure}

\begin{figure}
    \unitlength=1cm
    \begin{picture}(15,7)
        \unitlength=1mm
        \centerline{
        \epsfysize=7cm
        \epsfbox{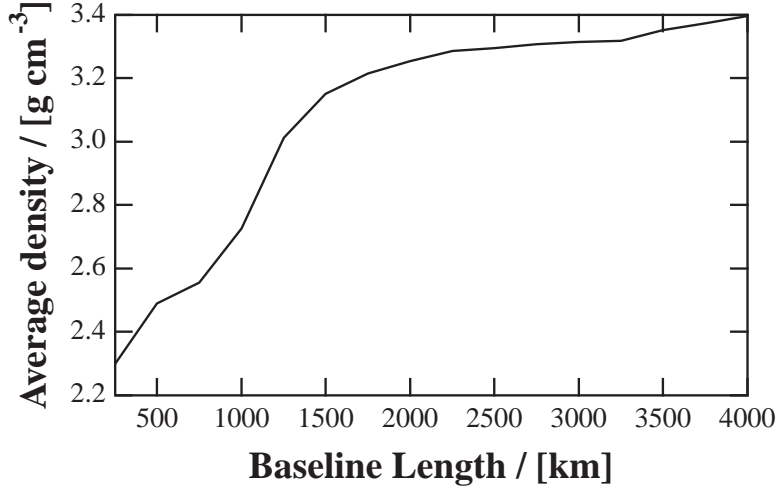}
        }
    \end{picture}
    \caption{%
    Approximated matter density as a function of baseline length,
    calculated from the Preliminary Reference Earth Model \cite{PREM}. 
    }
    \label{fig:density-PREM}
\end{figure}

We present example plots of ($N_{\mu} M_{\rm detector})_{\rm min; 1,
90\%}$ and ($N_{\mu} M_{\rm detector})_{\rm min; 2, 90\%}$ in
Figs.\ref{fig:chi1} and \ref{fig:chi2}.  
We adopt only single-bin method here because to divide the energy region
into some bins does not make the sensitivity better, especially in the
case without ambiguities of the parameters.
We will explain the reason in detail at Section\ref{subsec:binning}.
The parameters are taken as follows so that they are consistent to 
the present experimental limit:
\begin{equation}
    \sin\theta_{13} = 0.1, \quad
    \sin\theta_{23} = \frac{1}{\sqrt{2}}, \quad
    \sin\theta_{12} = 0.5,
    \label{eq:graph-angles}
\end{equation}
\begin{equation}
    \delta m^2_{31} = 3\times 10^{-3}{\rm eV}^2, \quad
    \delta m^2_{21} = 1\times 10^{-4}{\rm eV}^2.
    \label{eq:graph-m^{2}}
\end{equation}
CP-violating angle $\delta$ is taken to be $\pi/2$.  Matter effect
$a$ is approximated to be constant on the baseline, but its value
depends on the baseline length since the longer baseline gets deeper
in the earth.  The Preliminary Reference Earth Model \cite{PREM} is
adopted to estimate matter density as in 
Fig.\ref{fig:density-PREM} \cite{Ota}.
%
%

There is qualitative difference between Figs.\ref{fig:chi1} and
\ref{fig:chi2}.  Figure \ref{fig:chi1} shows that the sensitivity in
terms of $N(\delta) - N(\{ 0, \pi \})$ enhances as the muon energy
gets larger.  This does not hold for $\Delta N(\delta) - \Delta N(\{
0, \pi \})$ as seen in Fig.\ref{fig:chi2}.  There is a sweet spot in
this case in terms of muon energy and baseline length that optimizes
the sensitivity to the CP-violation effect.

We discuss the relation between $\chi^{2}_{1}$ and $\chi^{2}_{2}$.  In
the CP/T-violation search, one compare $( N_{j}(\delta), \bar
N_{j}(\delta) )$ with $( N_{j}(\delta_{0}), \bar N_{j}(\delta_{0}) )$. 
One can equivalently compare $( N^{\rm Total}_{j}(\delta), \Delta
N_{j}(\delta) )$ with $( N^{\rm Total}_{j}(\delta_{0}), \Delta
N_{j}(\delta_{0}) )$, where $N^{\rm Total}_{j}(\delta) \equiv 
N_{j}(\delta) + \bar N_{j}(\delta)$.  A $\chi^{2}$ statistics defined by
\begin{equation}
    \chi^{2}_{1'} (\delta_{0})
    \equiv
    \sum_{j = 1}^{n}
    \frac{
    [ N^{\rm Total}_{j}(\delta) - N^{\rm Total}_{j}(\delta_{0}) ]^{2}
    }{
    N^{\rm Total}_{j}(\delta)
    }
    +
    \sum_{j = 1}^{n}
    \frac{
    [ \Delta N_{j}(\delta) - \Delta N_{j}(\delta_{0}) ]^{2}
    }{
    N^{\rm Total}_{j}(\delta)
    }
    \label{eq:chi1-prime}
\end{equation}
obviously corresponds to $\chi^{2}_{1}(\delta_{0})$.  The second term
of eq.(\ref{eq:chi1-prime}) is $\chi^{2}_{2}(\delta_{0})$ itself. 
Hence we focus to the first term to understand the relation between
$\chi^{2}_{1}$ and $\chi^{2}_{2}$.  We note that $N^{\rm
Total}(\delta)$ is insensitive to CP-violation effect, since the
magnitude of genuine CP-violation effect for neutrinos and
antineutrinos are identical with opposite sign.  The term we are
discussing is thus insensitive to the imaginary coupling.  Our aim was
a direct CP/T-violation search or a search for an imaginary coupling,
and thus we dropped this term to obtain $\chi^{2}$.  On the other
hand, the dependence on the energy of $N^{\rm Total}$ and $\Delta N$
in high energy region is given by
\begin{eqnarray}
    N^{\rm Total} & \sim & E_{\mu},
    \label{eq:NTot-Edep} \\
    \Delta N & \sim & E_{\mu}^{0},
    \label{eq:DeltaN-Edep}
\end{eqnarray}
which follows from eqs.(\ref{eq:N-P-relation}),
(\ref{eq:hiE-paramFitProb}) and (\ref{eq:hiE-deltaProb}).  Hence the
first term gets larger as energy gets larger, and dominates the right
hand side of eq.(\ref{eq:DeltaN-Edep}) in the high energy region; total
fit gets better in spite of a poor fit of the imaginary coupling.  The
high sensitivity obtained by use of $\chi^{2}_{1}$ and shown in
Fig.\ref{fig:chi1} was achieved in this way.

The sweet spot seen in Fig.\ref{fig:chi2} can be intuitively
understood.  The CP-violation effect appears when the number of
generations is more than three \cite{KM}.  On one hand the heaviest state
decouples from the oscillation at the low energy region such that
$E_{\mu} \ll \delta m^{2}_{21} L$, and on the other hand the lighter
two generations are effectively degenerate in high energy region such
that $E_{\mu} > \delta m^{2}_{31} L$.  Thus the suitable energy
region to observe CP-violation effect is roughly given by $\delta
m^{2}_{21} L \lesssim E_{\mu} \lesssim \delta m^{2}_{31} L$.  The
sweet spot exactly lies in this region reflecting that
$\chi^{2}_{2}$ is indeed sensitive to the imaginary coupling.

\begin{figure}
 \unitlength = 1cm
 \begin{picture}(15,9)
  \unitlength = 1mm
  \centerline{
   \epsfysize = 9cm
   \epsfbox{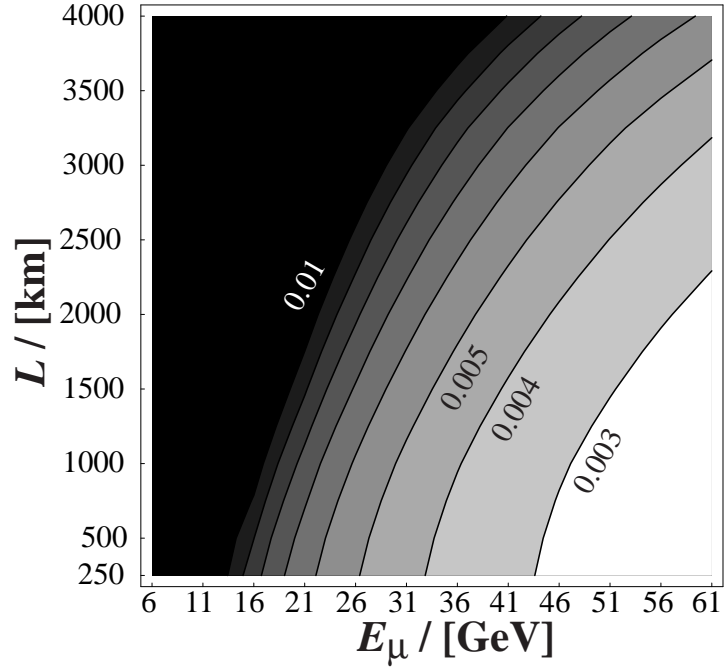}
  }
 \end{picture}
\caption{%
  Same as Fig.\ref{fig:chi1}, but here $\delta = - \pi / 2$.%
}
\label{fig:chi1-negative}
\end{figure}

\begin{figure}
 \unitlength = 1cm
 \begin{picture}(15,9)
  \unitlength = 1mm
  \centerline{
   \epsfysize = 9cm
   \epsfbox{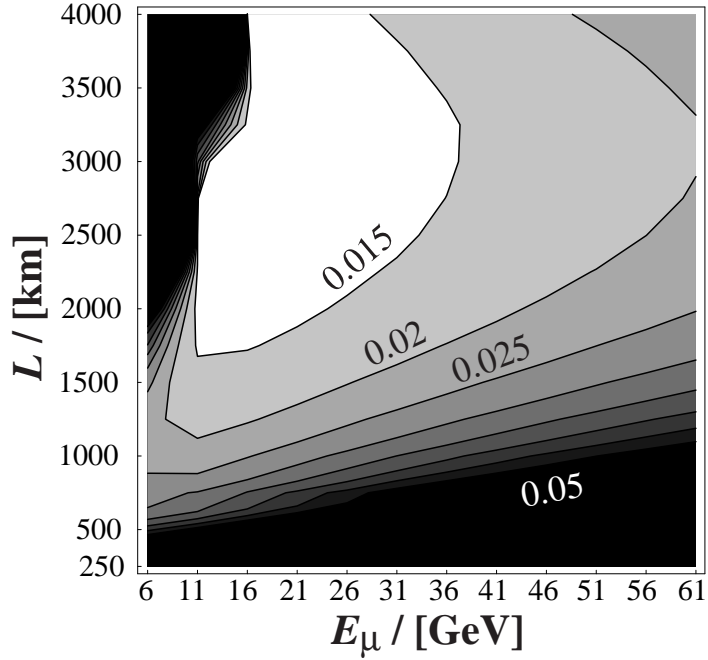}
  }
 \end{picture}
\caption{%
  Same as Fig.\ref{fig:chi2}, but here $\delta = - \pi / 2$.%
}
\label{fig:chi2-negative}
\end{figure}

We mention here that the sensitivity for the case $\delta = \pi / 2$
and for $\delta = - \pi / 2$ is not very different, contrary to the
discussion by other authors \cite{BGRW2}.  The previous works compared
$N(\delta)$ with $N(\delta_{0} = 0)$ alone, but we compared
$N(\delta)$ with both $N(\delta_{0} = 0)$ and $N(\delta_{0} = \pi)$. 
One should keep in mind that CP symmetry is conserved not only in the
$\delta = 0$ case but in the $\delta = \pi$ case; the imaginary
coupling is absent in both two cases.  The real coupling is different
for these two cases, and thus we need to distinguish an experimental
result with both of them.  We took these points into account by the
definition eqs.(\ref{eq:chi1-def}) and (\ref{eq:chi2-def}).  We
present in Figs.\ref{fig:chi1-negative} and \ref{fig:chi2-negative}
the sensitivity plot similar to Figs.\ref{fig:chi1} and
\ref{fig:chi2}, but this case for $\delta = -\pi / 2$.  We observe
indeed no qualitative difference between Figs.\ref{fig:chi1} and
\ref{fig:chi1-negative} and between Figs.\ref{fig:chi2} and
\ref{fig:chi2-negative}, respectively.

%
%
%

\begin{figure}
 \unitlength = 1cm
 \begin{picture}(15,9)
  \unitlength = 1mm
  \centerline{
   \epsfysize = 9cm
   \epsfbox{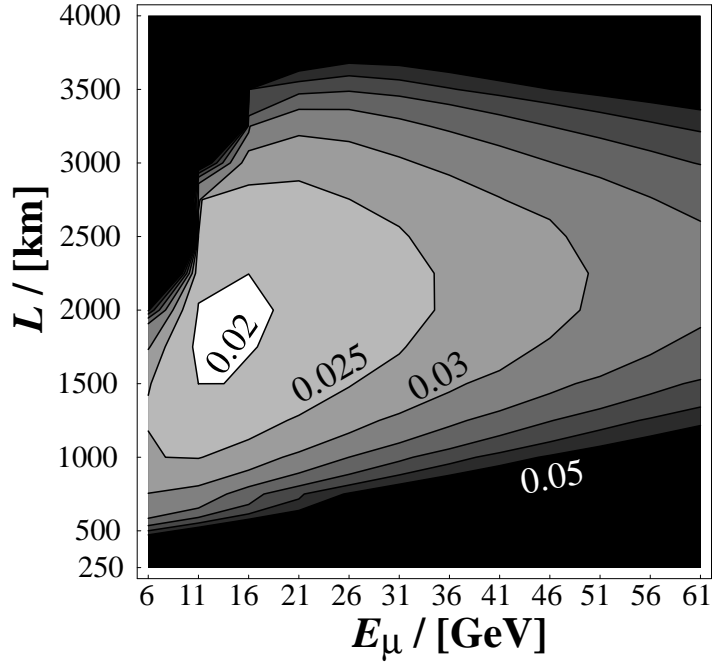}
  }
 \end{picture}
\caption{%
  A contour plot of $( N_{\mu} M_{\rm detector} )_{\rm min; 3, 90\%}$
  in unit of $\rm [10^{21} \cdot 100kt]$.  The parameters are the
  same as in Fig.\ref{fig:chi1}.  A sweet spot is seen in this figure
  as was also seen in Figs.\ref{fig:chi2} and \ref{fig:chi2-negative}.%
}
\label{fig:chi3}
\end{figure}

We have seen so far that we can extract the imaginary 
coupling by constructing the $\chi^{2}$ statistical quantity as in 
eq.(\ref{eq:chi2-def}).  The construction was motivated by taking the 
difference of event rates of neutrinos and antineutrinos.  We can build 
another quantity along this idea as
\begin{equation}
    N(\delta)
    -
    \frac{N(\delta_{0})}{\bar N(\delta_{0})} \bar N(\delta),
    \label{eq:chi3proto-1}
\end{equation}
which vanishes when $\delta = \delta_{0}$.  A $\chi^{2}$ statistics 
for this quantity is given by
\begin{equation}
    \chi^{2}_{3} (\delta_{0})
    =
    \sum_{j = 1}^{n}
    \frac{
     \left[
     \bar N(\delta_{0}) N(\delta) - N(\delta_{0}) \bar N(\delta)
     \right]^{2}
    }
    {
    \bar N(\delta_{0})^{2} N(\delta) + N(\delta_{0})^{2} \bar N(\delta)
    },
    \label{eq:chi3-def}
\end{equation}
The sensitivity condition with 90\% confidence level in $n$-bin method is
\begin{equation}
    \chi^{2}_{3} > \chi^{2}_{90\%}(n),
    \label{eq:chi3-cond}
\end{equation}
or
\begin{eqnarray}
    N_{\mu} M_{\rm detector}
    & > &
    ( N_{\mu} M_{\rm detector} )_{\rm min; 3, 90\%}
    \nonumber \\
    & \equiv &
    \frac{1}{C}
    \frac{E_{\mu}^{2}}{L^{2}}
    \frac{ \chi^{2}_{90\%}(n) }
    {
    \displaystyle
     \min_{\delta_{0} \in \{ 0, \pi \}}
     \left\{
     \sum_{j = 1}^{n}
     \frac{
     \left[
     \bar R(\delta_{0}) R(\delta) - R(\delta_{0}) \bar R(\delta)
     \right]^{2}
    }
    {
    \bar R(\delta_{0})^{2} R(\delta) + R(\delta_{0})^{2} \bar R(\delta)
    }
    \right\}
    }.
\end{eqnarray}
Figure \ref{fig:chi3} shows an example plot of $(N_{\mu} M_{\rm
detector})_{\rm min; 3, 90\%}$ for the parameters given by
eqs.(\ref{eq:graph-angles}) and (\ref{eq:graph-m^{2}}).  The graph is
similar to Figs.\ref{fig:chi2} and \ref{fig:chi2-negative}, which are
obtained from $\chi^{2}_{2}$.  The quantity $\chi^{2}_{3}$ is thus 
also sensitive to the imaginary coupling and suitable as a statistics 
for the direct CP-violation search.

%


\section{CP-violation search in presence of ambiguities of the 
parameters}

\subsection{Sensitivity to CP-violation effect in presence of 
ambiguities of the parameters}
\label{subsec:ambiguity-sensitivity}
We used $N(\delta_{0})$ and $\bar N(\delta_{0})$ in the definitions of
$\chi^{2}_{2}$ and $\chi^{2}_{3}$.  Exact values of mixing angles and
$\delta m^{2}$'s are required to obtain $N(\delta_{0})$ and $\bar
N(\delta_{0})$, but they are not known in practice.  We discuss how
the ambiguities of parameters spoil the sensitivity to the CP
violation.

The ambiguity of parameters is especially important when we make use
of CP conjugate oscillation channels.  It is because the genuine 
CP-violation effect in this case is contaminated by the matter effect. 
An estimation of matter effect is required, and the ambiguity of
parameters are obstacles to the estimation.  For the better
understanding, let us get back to $P(\nu_{\mu} \rightarrow \nu_{\rm
e}) - P(\bar \nu_{\mu} \rightarrow \bar \nu_{\rm e})$ in high energy
region given by eq.(\ref{eq:hiE-deltaProb}).  It consists of two
parts, matter effect part $\Delta P_{\rm Matter}$ and CP-violation
effect part $\Delta P_{\rm CPV}$:
\begin{eqnarray}
    & &
    \Delta P_{\rm Matter}
    \equiv
    2 \left(
    \frac{\delta m_{31}^{2} L}{4E}
    \right)^{3}
    \frac{a L}{4E}
    \nonumber \\
    & \times &
    \left\{
    \frac{2}{3} B \cos 2 \theta_{13}
    +
    \frac{\delta m^{2}_{21}}{\delta m^{2}_{31}}
    \left[
      \frac{1}{3}
      j \cos \delta 
      \left( 2 \cos 2 \theta_{13} - 1 \right)
      - 2 B \cos 2 \theta_{13} \sin^{2} \theta_{12}
      \right]
      \right\},
    \label{eq:Delta-P-Matter} \\
    & &
    \Delta P_{\rm CPV}
    \equiv
    -
    2 \left(
    \frac{\delta m_{31}^{2} L}{4E}
    \right)^{3}
    \frac{\delta m^{2}_{21}}{\delta m^{2}_{31}}
    j \sin \delta.
    \label{eq:Delta-P-CPV}
\end{eqnarray}
There is an ambiguity in $\Delta P_{\rm Matter}$ due to the
ambiguities in $\delta m^{2}$'s, $\theta$'s and $a$.  A sensitivity to
CP-violation part $\Delta P_{\rm CPV}$ is lost if the ambiguity of
$\Delta P_{\rm Matter}$ is larger than $\Delta P_{\rm CPV}$ itself. 
Ambiguity of all the parameters contribute to the ambiguity of $\Delta
P_{\rm Matter}$.  It is thus important to take into account
ambiguities of all the parameters.  It is expected that the ambiguity
of $\Delta P_{\rm Matter}$ is large when $\Delta P_{\rm Matter}$
itself is large.  Recalling that $\Delta P_{\rm Matter}$ is
proportional to baseline length $L$ (due to the factor $a L / (4 E)$),
one should obtain a poor sensitivity in the long baseline region.  It
is important to consider the sensitivity to the CP-violation effect
when the parameters are not precisely known.

We improve the discussion given in the previous section and formulate
how to take ambiguities of parameters into account in estimating the
sensitivity to the CP-violation effect.  Suppose that one uses the
parameters $\{ \tilde x_{i} \} \equiv \{ \tilde \theta_{12}, \tilde
\theta_{23}, \tilde \theta_{13}, \delta \tilde m^{2}_{21}, \delta
\tilde m^{2}_{31}, \tilde a \}$, which are different from the true
values $\{ x_{i} \} \equiv \{ \theta_{12}, \theta_{23}, \theta_{13},
\delta m^{2}_{21}, \delta m^{2}_{31}, a \}$, to calculate
$N_{j}(\delta_{0})$ and $\bar N_{j}(\delta_{0})$.  One will estimate
\begin{equation}
    \tilde N_{j}(\delta_{0})
    =
    N_{j}(
    \nu_{\alpha} \rightarrow \nu_{\beta};
    \{ \tilde x_{i} \}, \delta_{0}
    )
    \label{eq:tilde-Nj}
\end{equation}
instead of $N_{j}(\delta_{0})$, where $N_{j}( \nu_{\alpha} \rightarrow
\nu_{\beta}; \{ \tilde x_{i} \}, \delta_{0} )$ is evaluated from
eq.(\ref{eq:app-event-number}). Then matter effect can be overestimated
and hence the sensitivity to the CP violation can be spoiled.

First we see this using $\chi^2_2$. In this case one will estimate
the fake CP violation due to matter as follows,
\begin{equation}
    \Delta \tilde N_{j}(\delta_{0})
    =
    N_{j}(
    \nu_{\alpha} \rightarrow \nu_{\beta};
    \{ \tilde x_{i} \}, \delta_{0}
    )
    -
    N_{j}(
    \bar\nu_{\alpha} \rightarrow \bar\nu_{\beta};
    \{ \tilde x_{i} \}, \delta_{0}
    ),
    \label{eq:Delta-tilde-Nj}
\end{equation}
instead of $\Delta N_{j}(\delta_{0})$. We then obtain
\begin{equation}
    \tilde \chi^{2}_{2} (\delta_{0})
    \equiv
    \sum_{j=1}^{n}
    \frac
    { [ \Delta N_{j}(\delta) - \Delta \tilde N_{j}(\delta_{0}) ]^{2} }
    { N_{j}(\delta) + \bar N_{j}(\delta) }
    \label{eq:chi2tilde-0-def}
\end{equation}
instead of $\chi^{2}_{2}(\delta_{0})$ defined in
eq.(\ref{eq:chi2-0-def}).  Adjusting the parameters $\{ \tilde x_{i}
\}$ within the ambiguities, one can provide a better fit to the
expected values in no CP-violation case by minimizing $\chi^{2}$. 
One can nevertheless infer CP violation is present in 90\% confidence
level if one cannot make the value of $\chi^{2}$ smaller than
$\chi^{2}_{\rm 90\%}$.  We thus generalize eq.(\ref{eq:chi2-def}) and
define
\begin{equation}
    \tilde \chi_{2}^{2}
    \equiv
    \min_{ \delta_{0} \in \{0, \pi \}; \{ \tilde x_{i} \} }
    \tilde \chi^{2}_{2} (\delta_{0}).
\end{equation}
A criterion that CP-violation effect is observable for 90\% 
confidence level in $n$-bin method is given similar to 
eq.(\ref{eq:chi2-cond}) as
\begin{equation}
    \tilde \chi_{2}^{2} > \chi^{2}_{90\%} (n),
\end{equation}
which can be rewritten in terms of $N_{\mu} M_{\rm detector}$ as
\begin{eqnarray}
    N_{\mu} M_{\rm detector}
    & > &
    \left(
     N_{\mu} M_{\rm detector}
    \right)_{\rm min; 2, amb, 90\%}
    \nonumber \\
    & \equiv &
    \frac{1}{C}
    \frac{L^{2}}{E_{\mu}^{2}}
    \frac{ \chi^{2}_{90\%}(n) }
    {\displaystyle
    \min_{\delta_{0} \in \{ 0, \pi \}; \{ \tilde x_{i} \} }
    \left\{
    \sum_{j = 1}^{n}
    \frac
    {
    \left[
     \Delta R_{j}(\{ x_{i} \}, \delta) -
     \Delta R_{j}(\{ \tilde x_{i} \},\delta_{0})
    \right]^{2}
    }
    { R_{j}(\{ x_{i} \}, \delta) + \bar R_{j}(\{ x_{i} \}, \delta) }
    \right\}
    }.
    \label{eq:NM-min-CP2}
\end{eqnarray}

\begin{figure}
    \unitlength=1cm
    \begin{picture}(15,7)
        \unitlength=1mm
        \centerline{
        \epsfysize=7cm
        \epsfbox{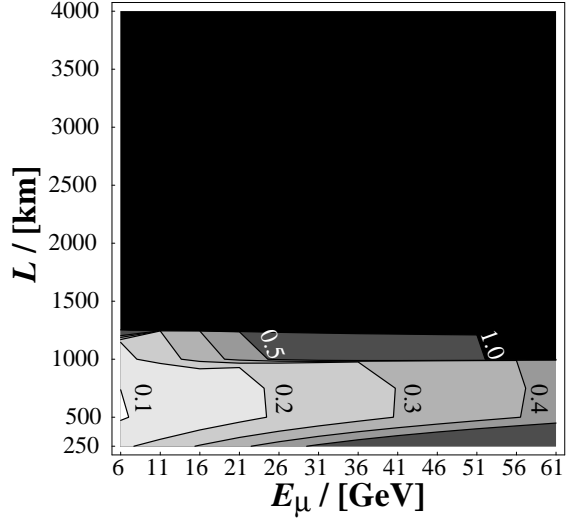}
        }
    \end{picture}
    \caption{%
    A contour plot of $(N_{\mu} M_{\rm detector})_{\rm min; 2, amb,
    90\%}$ in unit of $\rm [10^{21} \cdot 100kt]$.  The parameters are
    the same as in Fig.\ref{fig:chi2}.  The required data size is 
    much larger than the no-ambiguity case shown in 
    Fig.\ref{fig:chi2}.  The sensitivity is rapidly lost when the 
    baseline length gets longer than about one thousand kilometers.%
    }
  \label{fig:cp-ambiguity-chi2-1GeV}
\end{figure}
%
%
%

We present $(N_{\mu} M_{\rm detector})_{\rm min; 2, amb, 90\%}$ in
Fig.\ref{fig:cp-ambiguity-chi2-1GeV} 
to observe the CP-violation effect in 90\% confidence level.  All the
parameters $\{ x_{i} \}$ are assumed to have ambiguities of 10\%, and
their central value are taken as in eqs.(\ref{eq:graph-angles}),
(\ref{eq:graph-m^{2}}) and Fig.\ref{fig:density-PREM}, so that
\begin{eqnarray}
    0.09 < & \sin \tilde\theta_{13} & < 0.11,
    \nonumber \\
    \frac{0.9}{\sqrt{2}} <& \sin \tilde\theta_{23} &< \frac{1.1}{\sqrt{2}},
    \nonumber \\
    0.45 < & \sin \tilde\theta_{12} & < 0.55,
    \nonumber \\
    2.7 \times 10^{-3} {\rm eV}^{2}
    < & \delta \tilde m^2_{31} & <
    3.3 \times 10^{-3} {\rm eV}^2,
    \label{eq:param-range} \\
    0.9 \times 10^{-4}{\rm eV}^2
    < & \delta \tilde m^2_{21} & <
    1.1 \times 10^{-4}{\rm eV}^2,
    \nonumber \\
    0.9 a < & \tilde a & < 1.1 a.
    \nonumber
\end{eqnarray}
It is seen in both figures that genuine CP-violation effect is
difficult to be observed when the baseline length is longer than about
one thousand kilometers.  An estimation of matter effect is obscured
by the ambiguity when the baseline length is long, and CP-violation
effect cannot be separated from matter effect.  This result can be
understood qualitatively by the following rough estimation. 
CP-violation effect is hidden by the ambiguity of matter effect when
matter effect is large enough.  We require
\begin{equation}
    \left|
    \frac{\Delta P_{\rm CPV}}{\Delta P_{\rm Matter}}
    \right|
    \lesssim
    1
    \label{eq:P-ratio}
\end{equation}
as a rough estimation to observe CP-violation effect.%
\footnote{%
  One should actually use as a denominator the ambiguity of $\Delta
  P_{\rm Matter}$, not $\Delta P_{\rm Matter}$ itself.  We tentatively
  use eq.(\ref{eq:P-ratio}) to give a rough estimation, however.
}
Putting eqs.(\ref{eq:Delta-P-Matter}) and (\ref{eq:Delta-P-CPV})
into eq.(\ref{eq:P-ratio}), one obtain a condition on $L$ as
\begin{equation}
    L
    \lesssim
    \frac{4 E_{\nu}}{a}
    \frac{3 (\delta m^{2}_{21} / \delta m^{2}_{31}) j \sin \delta}
    { 2 \sin^{2} \theta_{23} \sin^{2} 2 \theta_{13} \cos 2 \theta_{13} }.
    \label{eq:L-condition}
\end{equation}
Applying our test parameters eqs.(\ref{eq:graph-angles}),
(\ref{eq:graph-m^{2}}) and Fig.\ref{fig:density-PREM} to
eq.(\ref{eq:L-condition}), one obtains
\begin{equation}
    L \lesssim 1250 {\rm km},
    \label{eq:L-condition2}
\end{equation}
which is consistent to Fig.\ref{fig:cp-ambiguity-chi2-1GeV}.

Next we illustrate using $\chi^2_3$ that the sensitivity to the CP
violation is lost in presence of ambiguities of parameters.  The
correspondent of eq.(\ref{eq:NM-min-CP2}) in this case is given by
\begin{eqnarray}
    &  &
    N_{\mu} M_{\rm detector}
    >
    ( N_{\mu} M_{\rm detector} )_{\rm min; 3, amb, 90\%}
    \nonumber \\
    & \equiv &
    \frac{1}{C}
    \frac{E_{\mu}^{2}}{L^{2}}
    \frac{ \chi^{2}_{90\%}(n) }
    {
    \displaystyle
    \min_{ \delta_{0} \in \{ 0, \pi \}; \{ \tilde x_{i} \} }
    \left\{
    \sum_{j = 1}^{n}
    \frac{
    \left[
    R_{j}(\{ \tilde x_{i} \}, \delta_{0})
    \bar R_{j}(\{ x_{i} \}, \delta) -
    \bar R_{j}(\{ \tilde x_{i} \}, \delta_{0})
    R_{j}(\{ x_{i} \}, \delta)
    \right]^2}
    {
    R_{j}(\{ \tilde x_{i} \}, \delta_{0})^{2}
    \bar R_{j}(\{ x_{i} \}, \delta) +
    \bar{R}_{j}(\{ \tilde x_{i} \}, \delta_{0})^{2}
    R_{j}(\{ x_{i} \}, \delta)
    }
    \right\}
    }.
    \label{eq:NM-min-CP3}
\end{eqnarray}
%
%
\begin{figure}
    \unitlength=1cm
    \begin{picture}(15,7)
        \unitlength=1mm
        \centerline{
        \epsfysize=7cm
        \epsfbox{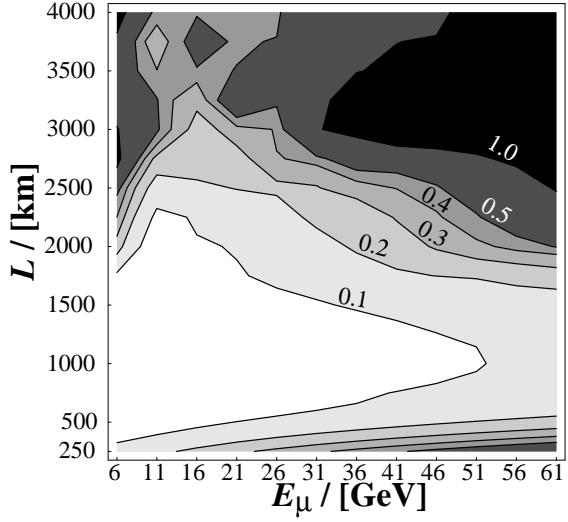}
        }
    \end{picture}
    \caption{%
    A contour plot of $( N_{\mu} M_{\rm detector} )_{\rm min; 3, amb,
    90\%}$ in unit of $\rm [10^{21} \cdot 100 kt]$.  The parameters
    are taken to be the same as Fig.\ref{fig:chi2}.  Note that the
    sensitivity is enhanced compared to
    Fig.\ref{fig:cp-ambiguity-chi2-1GeV}.%
    }
    \label{fig:cp-ambiguity-chi3-1GeV}
\end{figure}
We present a plot of $(N_{\mu} M_{\rm detector})_{\rm min; 3, amb, 90\%}$
in Fig.\ref{fig:cp-ambiguity-chi3-1GeV}.  The ambiguity of parameters
makes the sensitivity worse also in this case, as we see by comparing
Figs.\ref{fig:chi3} and \ref{fig:cp-ambiguity-chi3-1GeV}.  It is seen,
however, that the sensitivity shown in
Fig.\ref{fig:cp-ambiguity-chi3-1GeV} is better than that shown in
Fig.\ref{fig:cp-ambiguity-chi2-1GeV}, which means that $\chi^2_{3}$
avoids the ambiguity of matter effect better than $\chi^{2}_2$.  One
can understand the better sensitivity of $\chi^{2}_{3}$ as a
cancellation of ambiguity of $\sin \theta_{13}$ when the high energy
limit applies.  The dominant parts of $R$'s are given in the high 
limit by
%
\begin{eqnarray}
    R_j(\{x_i\},\delta) &=& B (S+T) + V
    \nonumber\\
    \bar R_j(\{x_i\},\delta) &=& B (S-T) - V
    \nonumber\\
    R_j(\{\tilde x_i\},\delta) &=& \tilde B (S+T)
    \label{eq:hiE-R-dominant}
    \\
    R_j(\{\tilde x_i\},\delta) &=& \tilde B (S-T),
    \nonumber
\end{eqnarray}
where
\begin{eqnarray}
    S
    &\equiv&
    \int_{E_{j-1}}^{E_j}
    E_{\nu}
    f_{\nu_{\alpha}} (E_{\nu})
    \left(
    \frac{\delta m_{31}^{2} L}{4E_\nu}
    \right)^{2}
    \frac{{\rm d} E_{\nu}}{E_{\mu}}
\nonumber\\
    T
    &\equiv&
\frac{2}{3}\cos 2\tilde\theta_{13}
    \int_{E_{j-1}}^{E_j}
    E_{\nu}
    f_{\nu_{\alpha}} (E_{\nu})
    \left(
    \frac{\delta m_{31}^{2} L}{4E_\nu}
    \right)^{3}
    \left(
    \frac{a L}{4E_\nu}
    \right)
    \frac{{\rm d} E_{\nu}}{E_{\mu}}
\\
    V
    &\equiv&
    j \sin\delta
    \int_{E_{j-1}}^{E_j}
    E_{\nu}
    f_{\nu_{\alpha}} (E_{\nu})
    \left(
    \frac{\delta m_{31}^{2} L}{4E_\nu}
    \right)^{3}
    \left(
    \frac{\delta m_{21}^{2}}{\delta m_{31}^{2}}
    \right)
    \frac{{\rm d} E_{\nu}}{E_{\mu}}.
    \\
\nonumber
\end{eqnarray}
Only the ambiguity of $\sin \theta_{13}$ is taken into account here,
and thus $\tilde B \equiv \sin \theta_{23} \sin^{2} 2
\tilde{\theta}_{13}$.  Using eqs.(\ref{eq:hiE-R-dominant}), we obtain
\begin{equation}
    \chi^{2}_{3}
    \simeq
    \frac{ ( \tilde B j \sin\delta )^{2} }{ 2 \tilde B^{2} B }
    =
    \frac{ ( j \sin\delta )^{2} }{ 2 B }.
    \label{eq:chi3-amb-cancel}
\end{equation}
Note here that $\tilde B$ vanishes in eq.(\ref{eq:chi3-amb-cancel}). 
The ambiguity of $\sin \theta_{13}$ is thus canceled away in the high
energy limit.  We also expect that the ambiguity does not spoil the
sensitivity to the CP-violation effect even in the lower energy.

\begin{figure}
    \unitlength=1cm
    \begin{picture}(15,17)
        \unitlength=1mm
        \centerline{
        \epsfysize=17cm
        \epsfbox{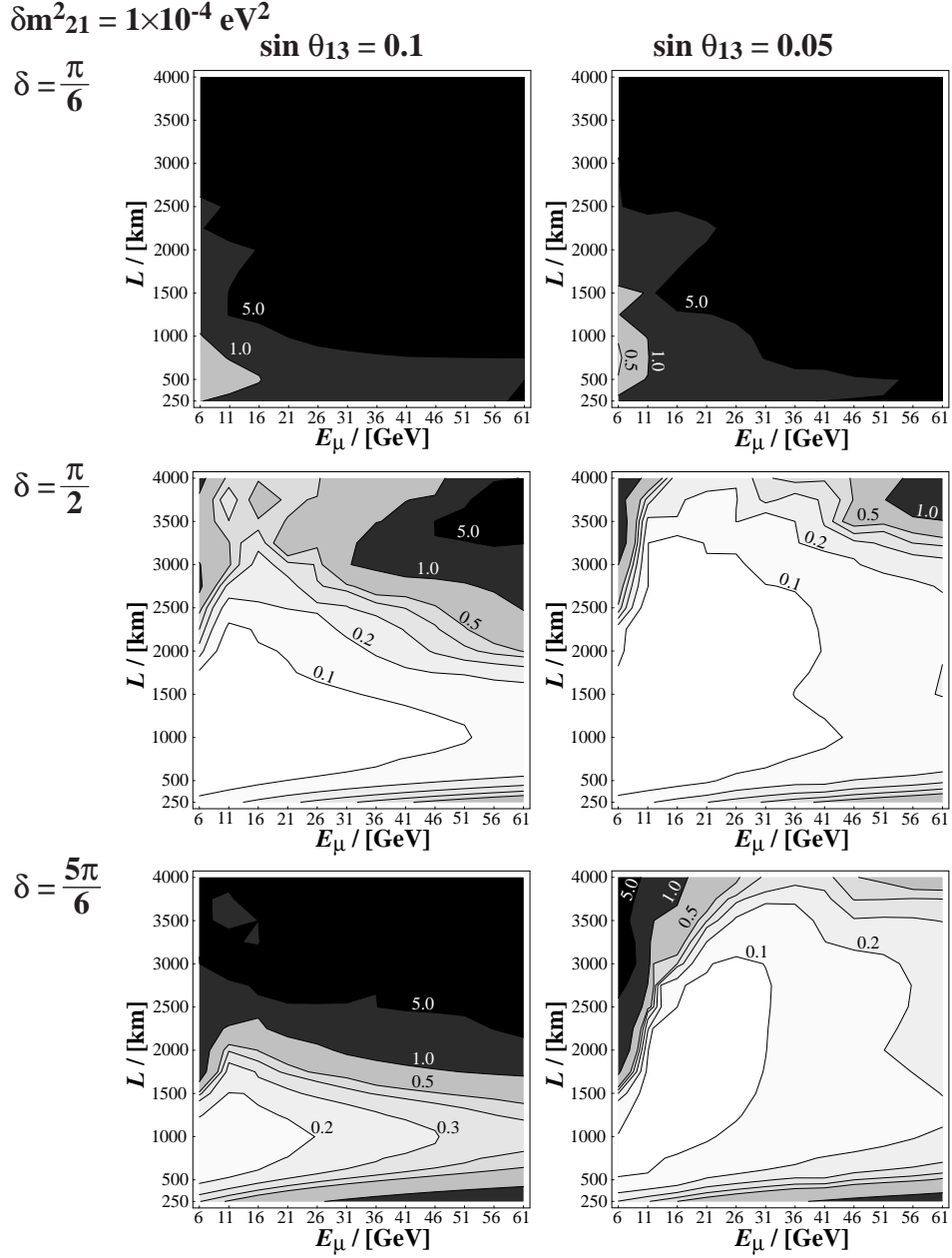}
        }
    \end{picture}
    \caption{%
     Same as Fig.\ref{fig:cp-ambiguity-chi3-1GeV}, but for different
     parameters.  All the graphs presented here are for
     $\delta m^{2}_{21} = 1 \times 10^{-4} {\rm eV}^{2}$.  The graphs in
     left column are for $\sin \theta_{13} = 0.1$ while the ones
     in right column are for $\sin \theta_{13} = 0.05$.  The top two
     graphs are for $\delta = \pi / 6$, the second two graphs are for
     $\delta = \pi / 2$, and the bottom two graphs are for $\delta = 5
     \pi / 2$.  Parameters not presented here are taken to be same as
     Fig.\ref{fig:chi2}.  The difference of the sensitivity for
     $\delta = \pi / 6$ and for $\delta = 5 \pi / 6$ is due to the
     difference of matter effect.%
     }
  \label{fig:1e-4-panel}
\end{figure}

\begin{figure}
    \unitlength=1cm
    \begin{picture}(15,20)
        \unitlength=1mm
        \centerline{
        \epsfysize=20cm
        \epsfbox{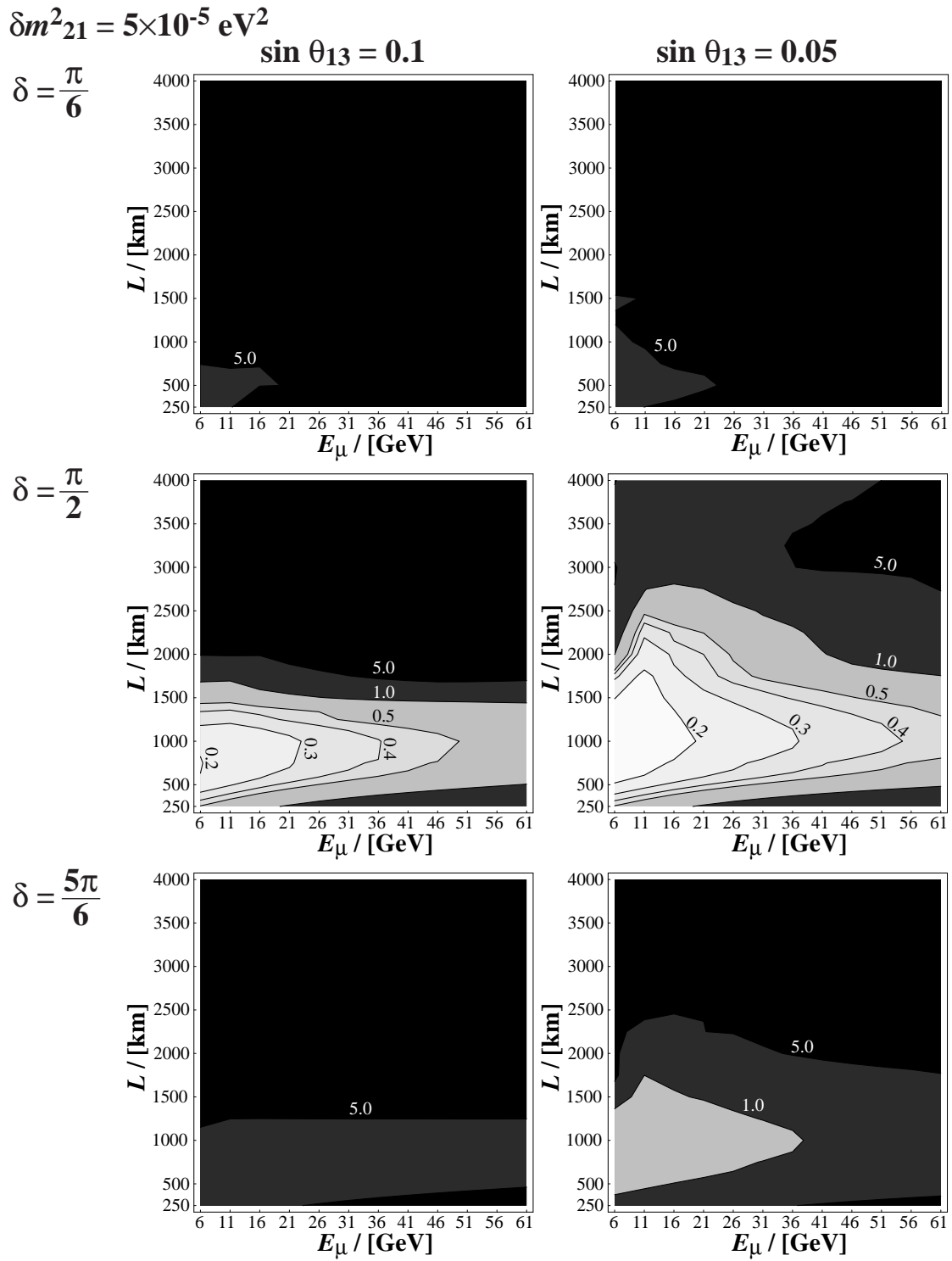}
        }
    \end{picture}
    \caption{%
    Same as Fig.\ref{fig:1e-4-panel}, but for $\delta m^{2}_{21} = 5 
    \times 10^{-5} {\rm eV}^{2}$.%
    }
  \label{fig:5e-5-panel}
\end{figure}

The sensitivity can be enhanced by a construction of a good statistics
such as $\chi^{2}_{3}$, but in general the sensitivity to the
imaginary part of coupling is smaller as the baseline length becomes
longer.  We confirm it for a couple of parameter sets by presenting
Figs.\ref{fig:1e-4-panel} and \ref{fig:5e-5-panel}.  It is also seen
in Fig.\ref{fig:1e-4-panel} that the sensitivity for $\delta = \pi
/ 6$ and for $\delta = 5\pi / 6$ is quite different.  The genuine
CP violation has a same magnitude for both of these two cases.  On the
other hand, the term proportional to $\cos \delta$, which is contained
in the matter effect term (see e.g. eq.(\ref{eq:Delta-P-Matter})), has
an opposite sign.  The magnitude of matter effect contamination is thus
different, and it leads to the difference of the sensitivity according
to our discussion that the sensitivity to CP-violation effect is
controlled by the magnitude of matter effect.
%
%

\begin{figure}
    \unitlength=1cm
    \begin{picture}(15,7)
        \unitlength=1mm
        \centerline{
        \epsfysize=7cm
        \epsfbox{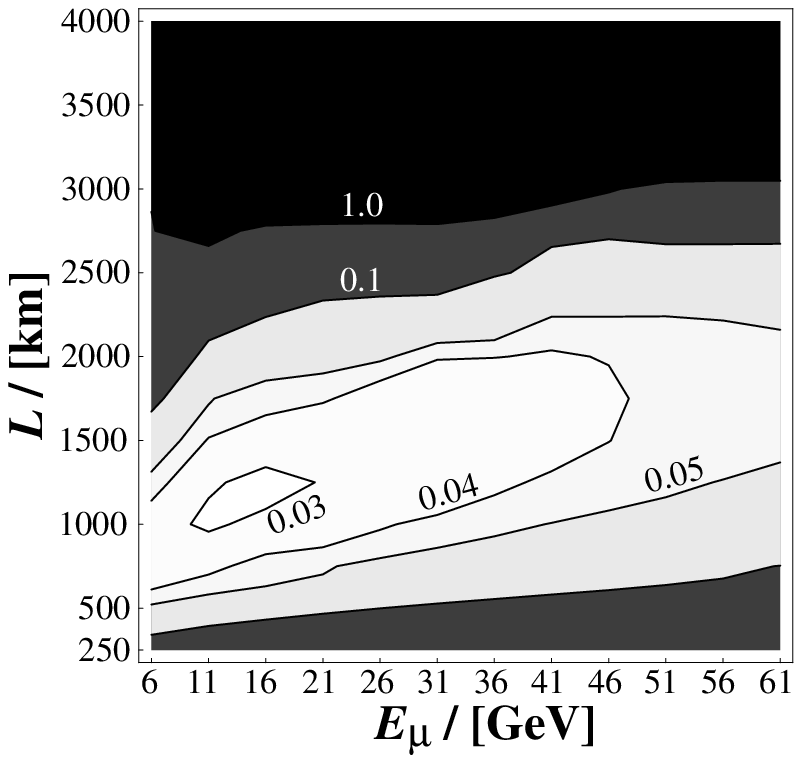}
        }
    \end{picture}
    \caption{%
    A contour plot for $(N_{\mu} M_{\rm detector})_{\rm min;
    1asym, amb, 90\%}$ defined by eq.(\ref{eq:delta-delta-cond}).  The
    sensitivity is lost also when the baseline length is longer than
    about 2,750 km.
    }
  \label{fig:delta-delta-ambiguity}
\end{figure}

We have been keeping on discussing the direct observation of
CP-violating effect.  One can verify our results by use of a quantity
similar to $\chi^{2}_{1}$ defined in eq.(\ref{eq:chi1-0-def}). 
Equation (\ref{eq:chi1-0-def}) focuses on the difference between
$N(\nu_{\alpha} \rightarrow \nu_{\beta}; \delta)$ and $N(\nu_{\alpha}
\rightarrow \nu_{\beta}; \delta_{0})$, and $\chi^{2}_{1}$ has little
sensitivity to the genuine CP-violation effect as a result.  Instead
we define
\begin{eqnarray}
    & &
    \chi_{\rm 1asym}^{2} (\delta_{0})
    \nonumber \\
    & \equiv &
    \sum_{j=1}^{n}
    \frac
    {[ N_{j}(\delta) - N_{j}(- \delta) ]^{2}}
    {N_{j}(\delta)}
    +
    \sum_{j=1}^{n}
    \frac
    {[ \bar N_{j}(\delta) - \bar N_{j}(- \delta) ]^{2}}
    {\bar N_{j}(\delta)}
    \nonumber \\
    & = &
    N_{\mu} M_{\rm detector}
    \frac{E_{\mu}^{2}}{L^{2}}
    C
    \left\{
    \sum_{j=1}^{n}
    \frac
    {[ R_{j}(\delta) - R_{j}(- \delta) ]^{2}}
    {R_{j}(\delta)}
    +
    \sum_{j=1}^{n}
    \frac
    {[ \bar R_{j}(\delta) - \bar R_{j}(- \delta) ]^{2}}
    {\bar R_{j}(\delta)}
    \right\},
    \label{eq:chi1asym-0-def}
\end{eqnarray}
which is sensitive to the genuine CP-violation
effect \cite{Lipari}.\footnote{%
  This quantity requires both CP- and T-conjugate channels; we
  consider this quantity just to verify the discussions so far.%
}
A quantity analogous to $(N_{\mu} M_{\rm detector})_{\rm min; 1, amb
90\%}$ is defined by
\begin{eqnarray}
    &&
    (N_{\mu} M_{\rm detector})_{\rm min; 1asym, amb, 90\%}
    \nonumber \\
    & \equiv &
    \frac{1}{C}
    \frac{L^{2}}{E_{\mu}^{2}}
    \frac{ \chi^{2}_{90\%}(2 n) }
    {\displaystyle
    \min_{ \{ \tilde{x}_{i} \}}
    \left\{
    \sum_{j=1}^{n}
    \frac
    {[ R_{j}(\{x_i\},\delta) - R_{j}(\{\tilde x_i\}, -\delta) ]^{2}}
    {R_{j}(\{x_i\},\delta)}
    +
    \sum_{j=1}^{n}
    \frac
    {[ \bar R_{j}(\{x_i\},\delta) - \bar R_{j}(\{\tilde x_i\},-\delta) ]^{2}}
    {\bar R_{j}(\{x_i\},\delta)}
    \right\}
    }.\nonumber \\ 
    \label{eq:delta-delta-cond}
\end{eqnarray}
Figure \ref{fig:delta-delta-ambiguity} shows a contour plot of
$(N_{\mu} M_{\rm detector})_{\rm min; 1asym, amb, 90\%}$.  It is seen
that the sensitivity is lost when the baseline length is longer than
about two thousand kilometers, which is qualitatively consistent to
the results obtained in this section.

%
%

\subsection{Energy dependence \label{subsec:binning}}

We discuss the binning of the neutrino energy in a search of CP 
violation.  We recall that the genuine CP violation in terms of 
oscillation probability is given by
\begin{equation}
    J
    \sin \frac{\delta m^{2}_{21} L}{4 E_{\nu}}
    \sin \frac{\delta m^{2}_{32} L}{4 E_{\nu}}
    \sin \frac{\delta m^{2}_{13} L}{4 E_{\nu}}.
    \label{eq:CPV-1}
\end{equation}
Applying $\delta m^{2}_{21} \ll \delta m^{2}_{31}$ and $\delta
m^{2}_{21} L / (4 E_{\nu}) \ll 1$, eq.(\ref{eq:CPV-1}) is rewritten to
be
\begin{equation}
    - J
    \frac{\delta m^{2}_{21} L}{4 E_{\nu}}
    \sin^{2} \frac{\delta m^{2}_{31} L}{4 E_{\nu}}.
    \label{eq:CPV-2}
\end{equation}
It can be seen from eq.(\ref{eq:CPV-2}) that the genuine CP-violation
effect has a definite sign as a function of $E_{\nu}$.  It is also
applicable to the event rate $N(\delta)$.  Dividing the event rates
into energy bins is thus meaningless and unnecessary to search for the
CP-violation effect, when the matter effect is absent.  All one need
to do is to observe the total counts of neutrinos.  This is of
practical importance for experimental studies since the determination of
neutrino energy is very challenging .
%
%
\begin{figure}
    \unitlength=1cm
    \begin{picture}(15,7)
        \unitlength=1mm
        \centerline{
        \epsfysize=7cm
        \epsfbox{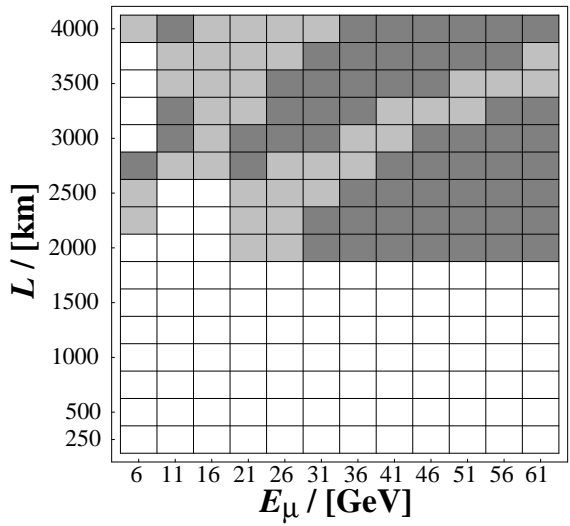}
        }
    \end{picture}
    \caption{%
    The most effective binning method to observe the CP-violation effect
    with $\chi_{3}^{2}$. 
    This corresponds to Fig.\ref{fig:cp-ambiguity-chi3-1GeV}.
    Single-, 3-, 5-bin analyses are compared.
    White, light gray, and dark gray regions mean where single-, 3-,
    5-bin analyses are most effective, respectively.
    This shows that multi-bin analysis is required in long baseline range.
    }
  \label{fig:binning-chi3}
\end{figure}

On the other hand, a single-bin analysis does not necessarily remain
advantageous once the matter effect is taken into account; the
multi-bin analysis is required to remove the matter in such a case
that the considered $\chi^{2}$ is sensitive not only to genuine
CP-violation effect but to matter effect (see Fig.\ref{fig:binning-chi3}).
The number of events per bin is sacrificed by the bin dividing, 
and each bin has a relatively small number of events compared to 
the single-bin analysis. 
As a result, the best-fit point of multi-bin analysis is less robust
than the single-bin analysis, i.e., the best fit point of multi-bin
analysis easily lies far away from the true parameter point.

We conclude as follows from the above considerations.  Experiments to
search for CP-violation effect should be made with the setup where the
single-bin analysis gives the best sensitivity, which means that the
matter effect contamination is small.  Such a setup has another
practical advantage in addition: measurements of the neutrino energy
are not required in single-bin analyses.  One need not take care of
the correlation between bins due to finite energy resolution, which
makes the sensitivity to the CP violation in the experiment worse.

\subsection{Dependence on $\sin \theta_{13}$ of the sensitivity to the
CP-violation effect}

We finally discuss the correlation between parameters on
the sensitivity to the direct CP-violation search.

The magnitude of CP violation is determined a single parameter, namely
the Jarlskog parameter $j \sin \delta$.  The correlation between
$\delta$ and other parameters such as $\theta_{13}$ is often
discussed, but it is heavily dependent on the parametrization.  The
presence or absence of CP/T violation can be determined without any
correlations to the mixing angles.%
\footnote{%
  The real part of coupling is in fact another intrinsic parameter
  which is independent of the parameterization.  The correlation
  between real part of the coupling and imaginary part of that will be
  present.  This is the only possible correlation for the CP-violation
  effect.
}%

\begin{figure}
    \unitlength=1cm
    \begin{picture}(15,7)
        \unitlength=1mm
        \centerline{
        \epsfysize=7cm
        \epsfbox{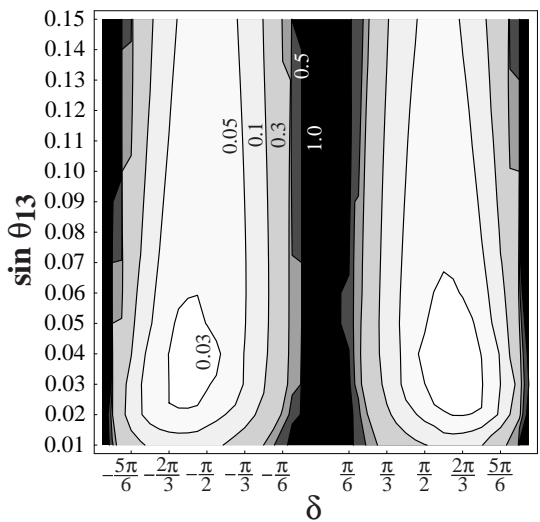}
        }
    \end{picture}
    \caption{%
    A similar plot to Fig.\ref{fig:cp-ambiguity-chi2-1GeV}, but this
    time $E_{\mu} = 10 {\rm GeV}$ and $L = 1000 {\rm km}$ are fixed
    while $\sin \theta_{13}$ and $\delta$ are varied.  The contour is
    nearly vertical, which reflects the fact that the value of $\sin
    \theta_{13}$ is not important to consider the sensitivity to
    CP-violation searches.%
    }%
    \label{fig:delta-vs-th31-1000km-10GeV}
\end{figure}
\begin{figure}
    \unitlength=1cm
    \begin{picture}(15,7)
        \unitlength=1mm
        \centerline{
        \epsfysize=7cm
        \epsfbox{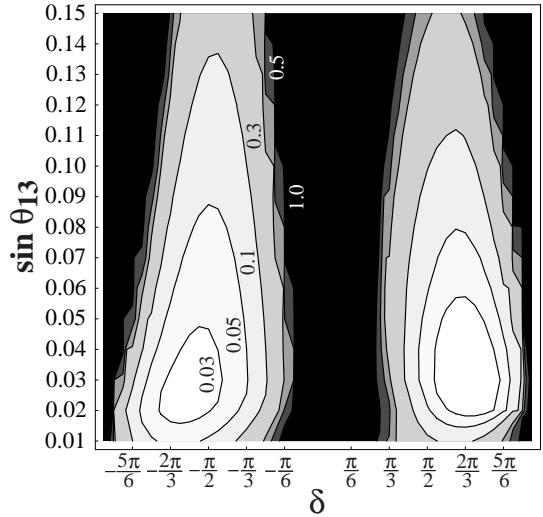}
        }
    \end{picture}
    \caption{%
    A similar plot to Fig.\ref{fig:delta-vs-th31-1000km-10GeV}, but 
    this time $E_{\mu} = 20 {\rm GeV}$ and $L = 2000 {\rm km}$.  
    Matter effect is larger in this parameter compared to 
    Fig.\ref{fig:delta-vs-th31-1000km-10GeV}, which leads to the 
    dependence of the sensitivity upon $\sin \theta_{13}$.%
    }%
    \label{fig:delta-vs-th31-2000km-20GeV}
\end{figure}

We present Figs.\ref{fig:delta-vs-th31-1000km-10GeV} and
\ref{fig:delta-vs-th31-2000km-20GeV} to show the correlation between
$\sin\theta_{13}$ and $\delta$.  They are sensitivity plots using
$\chi^{2}_{3}$, where $\delta$ and $\sin \theta_{13}$ are varied while
$E$ and $L$ are fixed.  Figure \ref{fig:delta-vs-th31-1000km-10GeV} is
a test plot for $E = 10 {\rm GeV}$ and $L = 1000 {\rm km}$, and
Fig.\ref{fig:delta-vs-th31-2000km-20GeV} is for $E = 20 {\rm GeV}$ and
$L = 2000 {\rm km}$.  A direct CP-violation search is expected to be
possible with this setup, as seen in Figs.\ref{fig:1e-4-panel} and 
\ref{fig:5e-5-panel}.
It is expected in these figures that the sensitivity scarcely
depends on $\theta_{13}$ if the statistics is correctly sensitive to
the genuine CP-violation effect.  It is illustrated by a rough
estimation of $\chi^{2}_{3}$ in such a case:
\begin{eqnarray}
    \chi_{3}^{2}
    & \sim &
    \frac{ (\Delta P_{\rm CPV})^{2} }
    { P(\nu_{\mu} \rightarrow  \nu_{\rm e}) }
    \nonumber \\
    & \sim &
    \frac{L^{2}}{E}
    \frac
    { \left[ (\delta m^{2}_{21} / \delta m^{2}_{31})
    j \sin \delta \right]^{2} }
    { B }
    \nonumber \\
    & = & 
    \frac{L^{2}}{E}
    \left( \frac{ \delta m^{2}_{21} }{ \delta m^{2}_{31} } \right)^{2}
    \frac{
      ( \sin 2 \theta_{12} \sin 2 \theta_{23}
      \sin 2 \theta_{13} \cos \theta_{13} )^{2}
    }{
      \sin^{2} \theta_{23} \sin^{2} 2 \theta_{13}
    }
    \nonumber \\
    & \sim &
    \frac{L^{2}}{E}
    \left( \frac{ \delta m^{2}_{21} }{ \delta m^{2}_{31} } \right)^{2}
    \sin^{2} 2 \theta_{12} \cos^{2} \theta_{23} \cos^{2} \theta_{13}.
    \label{eq:chi_{3}^{2}-estimation}
\end{eqnarray}
Equation (\ref{eq:chi_{3}^{2}-estimation}) depends on $\theta_{13}$
only through $\cos \theta_{13}$, which is almost unity for
$\theta_{13} \ll 1$.  The dependence on $\theta_{13}$ should be thus
quite small if the statistics is sensitive to $j \sin \delta$.  We
compare Fig.\ref{fig:delta-vs-th31-1000km-10GeV} with
Fig.\ref{fig:delta-vs-th31-2000km-20GeV} to confirm the above
discussion.  It can be seen in Fig.\ref{fig:cp-ambiguity-chi3-1GeV}
that the parameter for Fig.\ref{fig:delta-vs-th31-1000km-10GeV} is
more sensitive than that for Fig.\ref{fig:delta-vs-th31-2000km-20GeV}. 
We see that the dependence of the sensitivity upon $\sin \theta_{13}$
in Fig.\ref{fig:delta-vs-th31-1000km-10GeV} is quite small, while the
dependence is larger in Fig.\ref{fig:delta-vs-th31-2000km-20GeV}. 
This is indeed consistent to the above discussion; the larger matter
effect gives a sizable contribution to the numerator of $\chi^{2}_{3}$
in the latter case, and the estimation given in
eq.(\ref{eq:chi_{3}^{2}-estimation}) does not apply.  We thus conclude
that $L \sim 1000 {\rm km}$ and $E \sim 10 {\rm GeV}$ is the optimum
setup to search for a direct CP-violation search with use
of the statistics given by eq.(\ref{eq:chi3-def}).%
\footnote{%
  An optimum setup should change if one can find other better
  statistics, since the sensitivity itself depends on the adopted
  statistics.  The difference of Figs.\ref{fig:cp-ambiguity-chi2-1GeV}
  and \ref{fig:cp-ambiguity-chi3-1GeV} is an example.%
}%
%

%
%

%
\section{T-violation search in presence of ambiguities of parameters}

We have discussed in the previous section that the ambiguity of matter
effect spoils the sensitivity to the CP-violation effect.  One can
then expect that one can avoid the loss of sensitivity by use of
T-conjugate oscillation channels, which is free from the matter
effect \cite{HS}.

It is convenient to redefine
\begin{equation}
    N_{j}( \{ x_{i} \}, \delta )
    \equiv
    N(\nu_{\alpha} \rightarrow \nu_{\beta}; \{ x_{i} \}, \delta )
    \label{eq:Nj-def-T}
\end{equation}
and
\begin{equation}
    \bar N_{j}(\{ x_{i} \}, \delta)
    \equiv
    N(\nu_{\beta} \rightarrow \nu_{\alpha}; \{ x_{i} \}, \delta ),
    \label{eq:barNj-def-T}
\end{equation}
where $\alpha, \beta = {\rm e}, \mu, \tau$, and $\nu$ denotes
neutrinos and antineutrinos collectively.  We are to consider initial
neutrinos of different flavors, and we must take into account the
difference of the energy spectra of the neutrino beam.  For example,
the flux of $\nu_{\rm e}$ and $\nu_{\mu}$ obtained from the decay of
muons with energy $E_{\mu}$ is given in terms of $x \equiv E_{\nu} /
E_{\mu}$ by
\begin{eqnarray}
    f_{\nu_{\rm e}}(x) = 12 x^{2} (1 - x)
    \label{eq:nue-spectrum}
\end{eqnarray}
and
\begin{equation}
    f_{\nu_{\mu}}(x) = 2 x^{2} (3 - 2 x),
    \label{eq:numu-spectrum}
\end{equation}
respectively.  The quantity $\chi^{2}_{3}$ defined by
eq.(\ref{eq:chi3-def}) is suitable in such a case.  We define
$(N_{\mu} M_{\rm detector})_{\rm min; T, amb, 90\%}$ in the same
manner as the rightmost side of eq.(\ref{eq:NM-min-CP3}) so that
\begin{eqnarray}
    & &
    ( N_{\mu} M_{\rm detector} )_{\rm min; T, amb, 90\%}
    \nonumber \\
    & \equiv &
    \frac{1}{C}
    \frac{E_{\mu}^{2}}{L^{2}}
    \frac{ \chi^{2}_{90\%}(n) }
    {
    \displaystyle
    \min_{ \delta_{0} \in \{ 0, \pi \}; \{ \tilde x_{i} \} }
    \left\{
    \sum_{j = 1}^{n}
    \frac{
    \left[
    R_{j}(\{ \tilde x_{i} \}, \delta_{0})
    \bar R_{j}(\{ x_{i} \}, \delta) -
    \bar R_{j}(\{ \tilde x_{i} \}, \delta_{0})
    R_{j}(\{ x_{i} \}, \delta)
    \right]^2}
    {
    R_{j}(\{ \tilde x_{i} \}, \delta_{0})^{2}
    \bar R_{j}(\{ x_{i} \}, \delta) +
    \bar{R}_{j}(\{ \tilde x_{i} \}, \delta_{0})^{2}
    R_{j}(\{ x_{i} \}, \delta)
    }
    \right\}
    },
    \label{eq:NM-min-T}
\end{eqnarray}
while this time $R_{j}(\delta) = R_{j}(\nu_{\alpha} \rightarrow
\nu_{\beta}; \delta)$ and $\bar R_{j}(\delta) = R_{j}(\nu_{\beta}
\rightarrow \nu_{\alpha}; \delta)$.   T-violation effect is 
considered to be observable when
\begin{equation}
    N_{\mu} M_{\rm detector}
    >
    \left(
    N_{\mu} M_{\rm detector}
    \right)_{\rm min; T, amb, 90\%}
\end{equation}
is satisfied.

%
%
\begin{figure}
    \unitlength=1cm
    \begin{picture}(15,7)
        \unitlength=1mm
        \centerline{
        \epsfysize=7cm
        \epsfbox{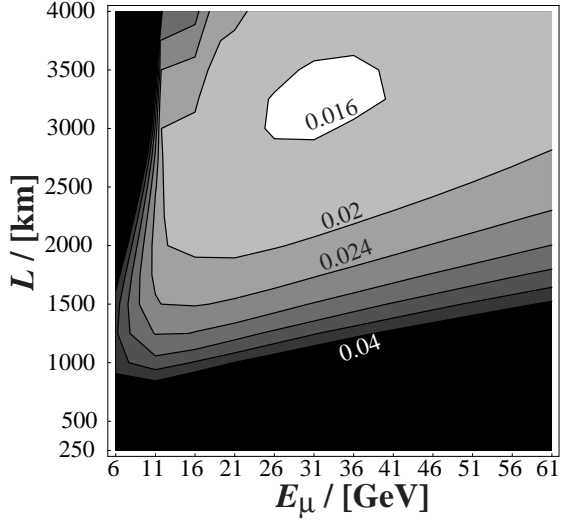}
        }
    \end{picture}
    \caption{%
    A contour plot of the required data size to observe the
    T-violation effect.  A quantity $(N_{\mu} M_{\rm detector})_{\rm
    min; T, amb, 90\%}$ defined in eq.(\ref{eq:NM-min-T}) is plotted
    in unit of $\rm [10^{21} \cdot 100kt]$.  The parameters are the
    same as in Fig.\ref{fig:chi2}.  This figure is similar to
    Fig.\ref{fig:chi3}, and it is seen that the use of T-conjugate
    channels is robust to the ambiguities of parameters.%
     }
  \label{fig:t-ambiguity}
\end{figure}

We present in Fig.\ref{fig:t-ambiguity} a test plot of $(N_{\mu}
M_{\rm detector})_{\rm min; T, amb, 90\%}$.  The parameters are taken as
shown in eq.(\ref{eq:param-range}).  It can be seen that
Fig.\ref{fig:t-ambiguity} is qualitatively similar to a plot in
absence of the ambiguity of parameters, which is presented in
Fig.\ref{fig:chi3}.  A sweet spot, which is expected from the naive
estimation in terms of oscillation probability, still remains in
Fig.\ref{fig:t-ambiguity}; thus we find that the CP/T-violation search
via T-conjugate channels is robust to the ambiguity of the parameters.

Longer baseline length is in general preferable when T-conjugate
channels are available, since eq.(\ref{eq:chi_{3}^{2}-estimation})
applies without being troubled with matter effect contamination.  This
is in contrast to the CP-conjugate case, where the matter effect
obscures the genuine CP-violation effect when the baseline gets too
long.



\section{Summary and Discussion}

We discussed the optimum experimental setup and the optimum analysis
to see CP-violation effect.

We first discussed the difference between the direct measurement and the
indirect measurement\cite{three}. Genuine CP-violation effect is
characterized by an imaginary part of couplings in Lagrangian and hence
quantities sensitive to this imaginary part should be used to measure
the CP violation.  To see this we introduced two statistical quantities,
$\chi^2_1$ (eq.(\ref{eq:chi1-def})) and $\chi_2^2$
(eq.(\ref{eq:chi2-def})).

Usually $\chi^2_1$ is used in analyses of neutrino factories.  We can
test using this whether the data can be explained by the hypothetical
data calculated assuming no CP-violation effect.  We saw, however, that
this quantity is not necessarily sensitive to the CP-violation part of
the coupling. In high energy region it is sensitive almost only to the
CP conserved part of the oscillation probability. We can tell about
genuine CP-violation effect only thorough unitarity relation of three
generation. The sensitivity to the CP-violation effect is often
indirect.  Thus we concluded that we should not use it to measure CP
violation since it often measures the CP-violation effect only
indirectly.

On the other hand, we can test with $\chi^2_2$ whether the asymmetry of
oscillation probabilities of neutrinos and antineutrinos exists.  We
have seen that $\chi^2_2$ is sensitive to the CP violating part of the
oscillation probability, and thus it is more suitable quantity to
measure the CP-violation effect. 

We saw the relation between $\chi^2_1$ and $\chi^2_2$ and found that to
pick up an imaginary part of couplings we need to see the difference
between particle and antiparticle. In this sense we also introduced a
statistics $\chi^2_3$ (eq.(\ref{eq:chi3-def})). This statistics gives
better sensitivity to measure CP-violation effect directly when we
consider the ambiguities of the theoretical parameters.

Then we investigated the influence of the ambiguities of the theoretical
parameters on $\chi^2_2$ and $\chi^2_3$.  Since the matter effect causes
between the difference the oscillation probabilities neutrinos and
antineutrinos, we have to estimate fake asymmetry to search for the 
CP-violation effect.  However, we will always ``overestimate'' the fake CP
violation because of the ambiguity of the theoretical parameters, and
hence we will always estimate the genuine CP-violation effect too small.
The matter effect increases as baseline length increases, and we will
lose the sensitivity to the asymmetry due to the genuine CP-violation
effect in longer baseline such as several thousand km.

The sensitivity of $\chi^2_2$ to genuine CP violation is lost much more
than that of $\chi^2_3$. This is due to the partial cancellation of
the ambiguity by $\sin\theta_{13}$. The ambiguity of estimation
of matter effect is partially canceled in the numerator.
We found that $\chi^2_3$ is better statistics to see CP-violation
effect directly.

Using $\chi^2_3$ we studied the correlation between $\sin\theta_{13}$
and $\delta$. Comparing Figures \ref{fig:delta-vs-th31-1000km-10GeV} and
\ref{fig:delta-vs-th31-2000km-20GeV}, we found that in general that we
have better sensitivity to CP violation with baseline length 1000km than
2000km. Moreover, if the statistics is only sensitive to the imaginary
part of couplings, Jarlskog parameter $J$, it has no dependence on
$\sin\theta_{13}$.\footnote{As long as $\sin\theta_{13}$ term has
dominant contribution. } Indeed in
Fig.\ref{fig:delta-vs-th31-1000km-10GeV} we see this behavior
while in Fig.\ref{fig:delta-vs-th31-2000km-20GeV}
we see strong dependence on $\sin\theta_{13}$ of the sensitivity.
In this sense we also understand that baseline length 1000km is better
to see CP violation directly.
Furthermore, Fig.\ref{fig:1e-4-panel} and Fig.\ref{fig:5e-5-panel} show 
that around 1000km is optimal baseline length for various parameter sets. 

Taking the statistics which is sensitive to the imaginary part of
the lepton couplings, we first showed that there is a sweet spot 
in terms of $E_{\mu}$ and $L$ when the ambiguities of the 
parameters are not considered.  We have then taken the ambiguities
of all the parameters to be 10\% and showed that the sweet spot 
survives in such a case.
We expect that the sweet spot also survives when we adopt the more
realistic values of the ambiguities.
We optimistically expect that other parameters will be determined 
with ambiguities less than 10\% except for $\theta_{13}$ in the
future.  The large ambiguity of $\theta_{13}$ is seemingly enough
to wash out the sweet spot.  We have mentioned in Section
\ref{subsec:ambiguity-sensitivity}, however, that the ambiguity of
$\theta_{13}$ is canceled by use of the statistics $\chi^{2}_{3}$.
We thus conclude that the experimental setup of $E_{\mu} \sim 10 {\rm 
GeV}$ and $L \sim 1000 {\rm km}$ is desirable even in the real 
experiment.

We finally studied T asymmetry.  There is no fake asymmetry due to
environmental effect such as the matter effect.  We found that the
naive expectation on CP-violation phenomena is indeed realized.

It is required to find another way to see CP-violation effect if we
can observe only appearance events of $\nu_{\rm e} \rightarrow
\nu_{\mu}$ and $\bar\nu_{\rm e} \rightarrow \bar\nu_{\mu}$.  Otherwise
we cannot observe CP-violation effect in neutrino factories with long
baseline ($\ge 1000$km) as the asymmetry between neutrinos and
antineutrinos. %
On the other hand, we can observe CP-violation effect as the T
asymmetry very well. Therefore it is very important to establish a way
to observe this asymmetry experimentally.

%

\appendix
\section*{Appendix: Statistics}
We explain a detail of the statistics used in this paper to estimate how
many events we need to tell the existence of the genuine CP-violation
effect. To state the feasibility of the experiment we consider not only
how well we can distinguish two theories (two parameter sets) but also
how well the best fit point lie in the true value. For example, even if
in nature $\delta=\pi/2$ is realized, we are not sure that the best fit
point for $\delta$ sit there. We will have the best fit point value
other than $\delta=\pi/2$ and hence we have to take care of this
possibility to state the feasibility of the experiment.

To estimate it, we employed the concept of
the ``power of test''. In the test, we
set up ``null hypothesis'', $H_{0}$, 
which should be rejected and its ``alternative hypothesis''
against $H_{0}$, $H_1$. In this paper we were interested in
whether we can insist the existence of the CP-violation effect,
and hence we set the null hypothesis,
\begin{eqnarray}
H_{0} \quad : \qquad \delta = \delta_{0}
\end{eqnarray}
and ``alternative hypothesis'' against $H_{0}$,
\begin{eqnarray}
H_{1} \quad : \qquad \delta \neq \delta_{0}.
\end{eqnarray}

We also define ``test statistics'' to give a criterion to reject
$H_0$ for a real data set $N^{\rm ex}_i$.
In this paper we examined three test statistics corresponding to
$\chi_{1}^{2}, \chi_{2}^{2} $, and $ \chi_{3}^{2}$:
\begin{align}
T_{1} (\delta_{0}) &\equiv \sum_{i}^{n}
         \frac{[ N^{\rm ex}_{i} - N^{{\rm th}}_{i}(\delta_{0}) ]^{2}}
              {N^{{\rm th}}_{i}(\delta_{0})}
         +
         \sum_{i}^{n}
         \frac{[\bar{N}^{\rm ex}_{i} - \bar{N}^{{\rm
               th}}_{i}(\delta_{0}) ]^{2}}
              {\bar{N}^{{\rm th}}_{i}(\delta_{0})},\quad 
   T_{1} \equiv \min_{\delta_{0} \in \{  0, \pi\} }
                T_{1} (\delta_{0}), \\
T_{2}(\delta_{0}) &\equiv \sum_{i}^{n}
         \frac{[ \{N^{\rm ex}_{i} - \bar{N}^{\rm ex}_{i}\}-
                 \{N^{{\rm th}}_{i}(\delta_{0}) 
                   - \bar{N}^{{\rm th}}_{i}(\delta_{0})\} ]^{2}}
              {N^{\rm th}_i(\delta_{0})+\bar{N}^{\rm th}_i(\delta_{0})},
   \quad
   T_{2} \equiv \min_{\delta_{0} \in \{  0, \pi\} }
                T_{2} (\delta_{0}),\\
T_{3}(\delta_{0}) &\equiv \sum_{i}^{n}
         \frac{[ \bar{N}^{{\rm th}}_{i}(\delta_{0}) \times N^{\rm ex}_{i}
                - N^{{\rm th}}_{i}(\delta_{0}) \times \bar{N}^{\rm ex}_{i}
               ]^{2}}
              {\{\bar{N}^{{\rm th}}_{i}(\delta_{0})\}^2 N^{\rm ex}_i
              +\{N^{{\rm th}}_{i}(\delta_{0})\}^2\bar{N}^{\rm ex}_i},
   \quad
   T_{3} \equiv \min_{\delta_{0} \in \{  0, \pi\} }
                T_{3} (\delta_{0}),
\end{align}
where $ N_{i}^{{\rm th}} $ is the event number assumed by the theory
with $ \delta $.
Hereafter we use as an example $\chi_1^{2}$ for the explanation. Furthermore
for simplicity we abbreviate $T_1$ (and accordingly $\chi_1^2$) as
\begin{align}
T_{1} &= \sum_{i}^{n}
         \frac{[ N^{\rm ex}_{i} - N^{{\rm th}}_{i}(\delta_{0}) ]^{2}}
              {N^{{\rm th}}_{i}(\delta_{0})}.
\end{align}

To reject $H_0$ at $\alpha$ ``level of significance,'' we require
\begin{eqnarray}
T_{1} > \chi^{2}_{\alpha} (2 n).
\label{eq:testineq}
\end{eqnarray}
Then the question is how well the inequality (\ref{eq:testineq})
is satisfied for given value $\delta$. This probability
is called ``power'';
\begin{eqnarray}
\beta_{1} (\delta) \equiv P_{\delta}(T_{1} > \chi^{2}_{\alpha}(2n)).
\label{eq:power}
\end{eqnarray}
This is the probability that we succeed in seeing the CP-violation
effect in the experiment. Thus we have to require  
that this probability should be larger than $\gamma$, 
which is almost 1.

To estimate the probability, often we generate
event sets with a given event rate and check whether $H_0$ is indeed
rejected according to the inequality (\ref{eq:testineq}) with
the probability $\gamma$.\footnote{We have to generate enough 
event sets enough to conclude that $H_0$ is rejected with the
probability $\gamma$.}
Instead to do so, here we make a following
approximation. First, we approximate $T_1$ as
\begin{align}
T_{1} &=    \sum_{i}^{n}
         \frac{[ N^{\rm ex}_{i} - N^{{\rm th}}_{i}(\delta_B) ]^{2}}
              {N^{{\rm th}}_{i}(\delta_B)}
     +
         \sum_{i}^{n}
     \frac{[ N^{\rm th}_{i}(\delta_B) - N^{{\rm th}}_{i}(\delta_{0}) ]^{2}}
              {N^{{\rm th}}_{i}(\delta_B)},
\label{eq:app1}
\end{align}
where $N^{{\rm th}}_{i}(\delta_B)$ is ``the maximum likelihood estimator'',
i.e.,
\begin{align}
\sum_{i}^{n}
         \frac{[ N^{\rm ex}_{i} - N^{{\rm th}}_{i}(\delta_B) ]^{2}}
              {N^{{\rm th}}_{i}(\delta_B)} \leq
& \sum_{i}^{n}
     \frac{[ N^{\rm ex}_{i} - N^{{\rm th}}_{i}(\delta) ]^{2}}
              {N^{{\rm th}}_{i}(\delta)},
\end{align}
for all $\delta$.
Equation (\ref{eq:app1}) holds well if 
$ | N^{\rm ex}_{i} - N^{{\rm th}}_{i}(\delta_B) | \leq O(\sqrt{
N^{{\rm th}}_{i}(\delta_B)})$, i.e., the fit for the data
$N^{\rm ex}_{i}$ by $N^{{\rm th}}_{i}(\delta_B)$ is good enough,
and $N^{{\rm th}}_{i}$ does not vary so rapidly around $\delta_B$.
We also assume that the estimator is almost the true value,
i.e., $\delta_B \simeq \delta$. Thus, 
\begin{align}
T_{1} &=    \sum_{i}^{n}
         \frac{[ N^{\rm ex}_{i} - N^{{\rm th}}_{i}(\delta) ]^{2}}
              {N^{{\rm th}}_{i}(\delta)} +
\sum_{i}^{n}
     \frac{[ N^{\rm th}_{i}(\delta) - N^{{\rm th}}_{i}(\delta_0) ]^{2}}
              {N^{{\rm th}}_{i}(\delta)} \nonumber \\
&=  \sum_{i}^{n}
         \frac{[ N^{\rm ex}_{i} - N^{{\rm th}}_{i}(\delta) ]^{2}}
              {N^{{\rm th}}_{i}(\delta)} +\chi_1^2 .
\end{align}

With this approximation we calculate the power (\ref{eq:power}) as follows:
\begin{eqnarray}
\beta_{1} (\delta) =
 P_{\delta}
  \left(
   \sum_{i}^{n}
         \frac{[ N^{\rm ex}_{i} - N^{{\rm th}}_{i}(\delta) ]^{2}}
              {N^{{\rm th}}_{i}(\delta)}
   >
   \chi^{2}_{\alpha}(2n) - \chi^{2}_{1}
  \right).
\end{eqnarray}
The left hand side in parenthesis of $P_{\delta}$ follows the $\chi^{2}$
distribution with $2n$ degrees of freedom so the requirement that the
power $\beta_{1} (\delta)$ should be larger than $\gamma$ is equivalent
to the condition
\begin{eqnarray}
\chi_{1}^{2} > \chi^{2}_{\alpha} (2n) - \chi^{2}_{\gamma} (2n-f),
\end{eqnarray}
where $f$ means the number of parameters included in the theory.
For example, if we take the 0.1 level of the significance and the
require the power to be 0.99 level, then
\begin{eqnarray}
\chi_{1}^{2} \geq \chi^{2}_{0.1} (2n) - \chi^{2}_{0.99} (2n-f).
\end{eqnarray}
Since in general if $\gamma\simeq 1$ then $\chi^{2}_{\gamma} (2n-f)$ is
very small for small $n$, it is omitted in this paper.\footnote{In other
words we required the perfect power, i.e., $\gamma=1$.} Thus we required%
\footnote{Since the significance level $\alpha$ corresponds naively to
the confidence level $(1-\alpha)\times 100$\%, we denote
$\chi^{2}_{\alpha} (n)$ by $\chi^{2}_{(1-\alpha)\times 100\%} (n)$ in
this paper.}
\begin{eqnarray}
\chi_{1}^{2} \geq \chi^{2}_{\alpha} (2n).
\end{eqnarray}

\section*{Acknowledgments}
The work of one of the authors (M.K.) is supported by the JSPS.
The work of J.S. is supported in part by a Grant-in-Aid for Scientific
Research of the Ministry of Education, Science and Culture,
No.12047221, and No.12740157.

%

\end{document}